\documentclass[secthm]{elsart}
\usepackage{amsmath,amssymb,color}
\usepackage{cite,graphicx,psfrag,url,setspace,lineno}

\newcommand{\N}{\mathbb{N}}
\newcommand{\R}{\mathbb{R}}

\newcommand{\lt}{\widetilde{\ell}}
\newcommand{\Spt}{\widetilde{S}_p}
\newcommand{\nnl}{\nonumber \\}
\newtheorem{lemma}[thm]{Lemma}

\newcommand{\RIP}{\text{RIP}}
\newcommand{\LQ}{\text{LQ}}
\newcommand{\hp}{\frac{2}{p}}
\newcommand{\ph}{\frac{p}{2}}
\renewcommand{\labelenumi}{(\roman{enumi})}

\begin{document}
\begin{frontmatter}
  \title{Sparse recovery by non-convex optimization -- instance
    optimality} \author{Rayan Saab} \address{Department of Electrical
    and Computer Engineering, University of British Columbia,
    Vancouver, B.C. Canada V6T 1Z4} \ead{rayans@ece.ubc.ca}
  \author[]{\"Ozg\"ur Y{\i}lmaz} \address{Department of Mathematics,
    University of British Columbia, Vancouver, B.C. Canada V6T 1Z2}
  \ead{oyilmaz@math.ubc.ca} \thanks{This work was supported in
    part by a Discovery Grant and by a CRD Grant (DNOISE) from the
    Natural Sciences and Engineering Research Council of
    Canada. R. Saab also acknowledges a UGF award from the
    UBC, and a Pacific Century Graduate
    Scholarship from the Province of British Columbia through the
    Ministry of Advanced Education. 
}

\begin{abstract}
  In this note, we address the theoretical properties of $\Delta_p$, a
  class of compressed sensing decoders that rely on $\ell^p$
  minimization with $0<p<1$ to recover estimates of sparse and
  compressible signals from incomplete and inaccurate measurements. In
  particular, we extend the results of Cand{\`e}s, Romberg and Tao
  \cite{CRT05} and Wojtaszczyk \cite{Wojtaszczyk08} regarding the
  decoder $\Delta_1$, based on $\ell^1$ minimization, to $\Delta_p$
  with $0<p<1$. Our results are two-fold. First, we show that
  under certain sufficient conditions that are weaker than the
  analogous sufficient conditions for $\Delta_1$ the decoders
  $\Delta_p$ are robust to noise and stable in the sense that they are
  $(2,p)$ instance optimal for a large class of encoders. Second, we
  extend the results of Wojtaszczyk to show that, like $\Delta_1$, the
  decoders $\Delta_p$ are $(2,2)$ instance optimal in probability
  provided the measurement matrix is drawn from an appropriate
  distribution.  
\end{abstract}
%\maketitle
\end{frontmatter}

\section{Introduction}
The sparse recovery problem received a lot of attention lately, both
because of its role in transform coding with redundant dictionaries
(e.g., \cite{vandergheynst2001eir,iventura2004lra,chen99atomic}), and
perhaps more importantly because it inspired compressed sensing
\cite{Donoho2006_CS,CRT06,CRT05}, a novel method of acquiring signals
with certain properties more efficiently compared to the classical
approach based on Nyquist-Shannon sampling theory.
Define $\Sigma_S^N$ to be the set of all \emph{$S$-sparse vectors}, i.e.,
$$ \Sigma_S^N := \{x \in \R^N:\ \ |\text{supp}(x)| \leq S\},$$
and define \emph{compressible} vectors as vectors that can be well
approximated in $\Sigma_S^N$.
Let $\sigma_S(x)_{\ell^p}$ denote the best $S$-term approximation
error of $x$ in $\ell^p \quad$ (quasi-)norm where $p>0$, i.e., 
$$\sigma_S(x)_{\ell^p} := \min_{v\in \Sigma_S^N} \|x-v\|_p.$$
Throughout the text, $A$ denotes an $M \times N$ real matrix where
$M<N$. Let the associated {\em{encoder}} be the map $x \mapsto Ax$
(also denoted by $A$).
The transform coding and compressed sensing problems mentioned above
require the existence of decoders, say $\Delta: \ \R^M \mapsto \R^N$,
with roughly the following properties:
\begin{enumerate}
\item[(C1)] $\Delta(Ax) = x$ whenever $x \in \Sigma_S^N$ with
  sufficiently small $S$.

\item[(C2)] $\|x-\Delta(Ax+e)\| \lesssim \|e\| +
  \sigma_S(x)_{\ell^p}$, where the norms are appropriately
  chosen. Here $e$ denotes measurement error, e.g., thermal and
  computational noise.

\item[(C3)] $\Delta(Ax)$ can be computed efficiently (in some sense).
\end{enumerate}
Below, we denote the (in general noisy) encoding of $x$ by $b$, i.e.,
\begin{equation} \label{eq:problem}
b=Ax+e.
\end{equation}
In general, the problem of constructing decoders with properties
(C1)-(C3) is non-trivial (even in the noise-free case) as $A$ is
overcomplete, i.e., the linear system of $M$ equations in
\eqref{eq:problem} is underdetermined, and thus, if consistent, it
admits infinitely many solutions. In order for a decoder to satisfy
(C1)-(C3), it must choose the ``correct solution" among these
infinitely many solutions.  Under the assumption that the original
signal $x$ is sparse, one can phrase the problem of finding the
desired solution as an optimization problem where the objective is to
maximize an appropriate ``measure of sparsity'' while simultaneously
satisfying the constraints defined by \eqref{eq:problem}. In the
noise-free case, i.e., when $e=0$ in \eqref{eq:problem}, under certain
conditions on the $M\times N$ matrix $A$, i.e., if $A$ is in general
position, there is a decoder $\Delta_0$ which
satisfies $\Delta_0(Ax)=x$ for all $x \in \Sigma_S^N$ whenever
$S<M/2$, e.g., see \cite{Donoho03}. This $\Delta_0$ can be explicitly
computed via the optimization problem
\begin{equation}
\Delta_0(b):=\arg \min_y \|y\|_0 \text{ subject to }
b=Ay \label{eq:problem_l0}.
\end{equation}
Here $\|y\|_0$ denotes the number of non-zero entries of the vector
$y$, equivalently its so-called $\ell^0$-norm. Clearly, the sparsity
of $y$ is reflected by its $\ell^0$-norm. 

\subsection{Decoding by $\ell^1$ minimization}\label{sec:Intro_A}
As mentioned above, $\Delta_0(Ax)=x$ exactly if $x$ is sufficiently
sparse depending on the matrix $A$. However, the associated
optimization problem is combinatorial in nature, thus its complexity
grows quickly as $N$ becomes much larger than $M$. Naturally, one then
seeks to modify the optimization problem so that it lends itself to
solution methods that are more tractable than combinatorial search. In
fact, in the noise-free setting, the decoder
defined by $\ell^1$ minimization, given by
\begin{equation}
\Delta_1(b):=\arg\min_y\|y\|_1 \text{ subject to } Ay=b, \label{eq:problem_l1}
\end{equation}
recovers $x$ exactly if $x$ is sufficiently sparse and the matrix $A$
has certain properties (e.g.,
\cite{CandesTao05,CRT05,Donoho03,Donoho01,chen99atomic,tropp05}).
In particular, it has been shown in \cite{CRT05} that if
$x\in\Sigma_S^N$ and $A$ satisfies a certain \emph{restricted isometry
  property}, e.g., $\delta_{3S}<1/3$ or more generally
$\delta_{(k+1)S}<\frac{k-1}{k+1}$ for some $k>1$ such that $k\in
\frac{1}{S}\N$, then $\Delta_1(Ax)=x$ (in what follows, $\N$ denotes
the set of positive integers, i.e., $0\notin \N$). Here $\delta_{S}$
are the \emph{$S$-restricted isometry constants} of $A$, as introduced
by Cand{\`e}s, Romberg and Tao (see, e.g., \cite{CRT05}), defined as
the smallest constants satisfying
\begin{equation}\label{eq:UUP}
(1-\delta_S)\|c\|_2^2 \leq \|A c\|_2^2 \leq
(1+\delta_S)\|c\|_2^2
\end{equation}
for every $c \in \Sigma_S^N$. Throughout the paper, using the
notation of \cite{Wojtaszczyk08}, we say that a matrix satisfies
$\RIP(S,\delta)$ if $\delta_S<\delta$. 

Checking whether a given matrix satisfies a certain RIP is
computationally intensive, and becomes rapidly intractable as the size
of the matrix increases. On the other hand, there are certain classes
of random matrices which have favorable RIP. In fact, let $A$ be an $M
\times N$ matrix the columns of which are independent, identically
distributed (i.i.d.) random vectors with any sub-Gaussian
distribution. It has been shown that $A$ satisfies $\RIP\left(S,
 \delta\right)$ with any $0<\delta<1$ when
\begin{equation}S \leq c_1M/log(N/M) ,\end{equation} with probability
greater than
$1-2e^{-c_2M}$ (see, e.g.,
\cite{baraniuk100spr},\cite{CandesTao05_2},\cite{CandesTao05}), where
$c_1$ and $c_2$ are positive constants that only depend on $\delta$
and on the actual distribution from which $A$ is drawn.

In addition to recovering sparse vectors from error-free observations,
it is important that the decoder be robust to noise and stable with
regards to the ``compressibility" of $x$. In other words, we require
that the reconstruction error scale well with the measurement error
and with the ``non-sparsity'' of the signal (i.e., (C2) above). For
matrices that satisfy $\RIP((k+1)S,\delta)$, with
$\delta<\frac{k-1}{k+1}$ for some $k>1$ such that $k\in\frac{1}{S}\N$,
it has been shown in \cite{CRT05} that there exists a feasible decoder
$\Delta_1^\epsilon$ for which the approximation error
$\|\Delta_1^\epsilon(b)-x \|_2$ scales linearly with the measurement
error $\|e\|_2\leq\epsilon$ and with $\sigma_S(x)_{\ell^1}$.  More
specifically, define the decoder
\begin{equation}
  \Delta_1^\epsilon(b)=\arg\min_y\|y\|_1 \text{ subject to } \|Ay-b\|_2 \leq\epsilon \label{eq:problem_l1_eps}.
\end{equation}
The following theorem of Cand{\`e}s et al. in \cite{CRT05} provides
error guarantees when $x$ is not sparse and when the observation is
noisy.

\begin{thm} \label{thm:P2} {\em\cite{CRT05}} Fix $\epsilon \ge 0$,
  suppose that ${x}$ is arbitrary, and let ${b}={A}{x} + {e}$ where
  $\|{e}\|_2\leq\epsilon$. If $A$ satisfies \mbox{$\delta_{3S}
    +3\delta_{4S} < 2$}, then
\begin{equation} \|\Delta^\epsilon_1(b)-{x} \|_2 \leq
C_{1,S}\epsilon + C_{2,S} \frac{\sigma_S(x)_{\ell^1}}{\sqrt{S}}.
\label{delta_1_error_compressible} 
\end{equation} 
For reasonable values of $\delta_{4S}$, the constants are well
behaved; e.g., $C_{1,S} = 12.04$ and $C_{2,S}=8.77$ for $\delta_{4S} =
1/5$.
\end{thm}

\begin{rem}{\normalfont This means that given $b=Ax+e$, and $x$ is
sufficiently sparse, $\Delta_1^\epsilon(b)$ recovers the underlying
sparse signal within the noise level. Consequently the recovery is
perfect if $\epsilon=0$. }\end{rem}

\begin{rem}\label{thm:P1}{\normalfont By explicitly assuming $x$ to be
sparse, Cand{\`e}s et. al. \cite{CRT05} proved a version of the above
result with smaller constants, i.e., for $b=Ax+e$ with $x \in \Sigma_S^N$
and $\|e\|_2 \le \epsilon$,
\begin{equation} 
\|\Delta^\epsilon_1(b)-{x} \|_2 \leq C_S \epsilon\label{delta_1_error},
\end{equation}
where $C_S<C_{1,S}$.}
\end{rem}

\begin{rem}{\normalfont Recently, Cand{\`e}s \cite{candes2008rip}
    showed that $\delta_{2S}< \sqrt{2}-1$ is sufficient to guarantee
    robust and stable recovery in the sense of
    \eqref{delta_1_error_compressible} with slightly better
    constants.}
\end{rem}

In the noise free case, i.e., when $\epsilon=0$, the reconstruction
error in Theorem \ref{thm:P2} is bounded above by
$\sigma_S(x)_{\ell^1}/\sqrt{S}$, see
\eqref{delta_1_error_compressible}. This upper bound would sharpen if
one could replace $\sigma_{S}(x)_{\ell^1}/\sqrt{S}$ with
$\sigma_{S}(x)_{\ell^2}$ on the right hand side of
\eqref{delta_1_error_compressible} (note that $\sigma_S(x)_{\ell^1}$
can be large even if all the entries of the reconstruction error are
small but nonzero; this follows from the fact that for any vector $y
\in \R^N$, $\|y\|_2\le \|y\|_1 \le \sqrt{N} \|y\|_2$, and consequently
there are vectors $x\in \R^N$ for which $\sigma_S(x)_{\ell^1}/\sqrt{S}\gg
\sigma_S(x)_{\ell^2}$, especially when $N$ is large). In
\cite{cohen2006csa} it was shown that the term
$C_{2,S}\sigma_S(x)_{\ell^1}/\sqrt{S}$ on the right hand side of
\eqref{delta_1_error_compressible} {\em cannot} be replaced with
$C\sigma_S(x)_{\ell^2}$ if one seeks the inequality to hold for all $x
\in \R^N$ with a fixed matrix $A$, unless $M> cN$ for some constant $c$. This is
unsatisfactory since the paradigm of compressed sensing relies on the
ability of recovering sparse or compressible vectors $x$ from
significantly fewer measurements than the ambient dimension $N$.

Even though one cannot obtain bounds on the approximation error in
terms of $\sigma_S(x)_{\ell^2}$ with constants that are uniform on $x$
(with a fixed matrix $A$), the situation is significantly better if we
relax the uniformity requirement and seek for a version of
\eqref{delta_1_error_compressible} that holds ``with high
probability''. Indeed, it has been recently shown by Wojtaszczyk that
for any specific $x$, $\sigma_S(x)_{\ell^2}$ can be placed in
\eqref{delta_1_error_compressible} in lieu of
$\sigma_S(x)_{\ell^1}/\sqrt{S}$ (with different constants that
are still independent of $x$) with
high probability on the draw of $A$ if (i) $M >c S\log{N}$ and (ii)
the entries ${A}$ is drawn i.i.d. from a Gaussian distribution
or the columns of $A$ are drawn i.i.d. from the uniform
distribution on the unit sphere in $\R^M$ \cite{Wojtaszczyk08}. In
other words, the encoder $\Delta_1=\Delta_1^0$ is ``(2,2) instance
 optimal in probability'' for encoders associated with such $A$, a
property which was discussed in \cite{cohen2006csa}.

Following the notation of \cite{Wojtaszczyk08}, we say that an
encoder-decoder pair $(A,\Delta)$ is \emph{$(q,p)$ instance optimal of
  order $S$ with constant C} if
\begin{equation}
\|\Delta({Ax})-x\|_q \leq C\frac{\sigma_S(x)_{\ell^p}}{S^{1/p-1/q}}
\label{eq:IO}
\end{equation}
holds for all $x \in \R^N$.  Moreover, for random matrices $A_\omega$,
$(A_\omega,\Delta)$ is said to be \emph{$(q,p)$ instance optimal in
  probability} if for any $x$ \eqref{eq:IO} holds with high
probability on the draw of $A_\omega$.  Note that with this notation
Theorem \ref{thm:P2} implies that $(A,\Delta_1)$ is (2,1) instance
optimal (set $\epsilon=0$), provided $A$ satisfies the conditions of
the theorem.

The preceding discussion makes it clear that $\Delta_1$ satisfies
conditions (C1) and (C2), at least when $A$ is a sub-Gaussian random
matrix and $S$ is sufficiently small. It only remains to note that
decoding by $\Delta_1$ amounts to solving an $\ell^1$ minimization
problem, and is thus tractable, i.e., we also have (C3). In fact,
$\ell^1$ minimization problems as described above can be solved
efficiently with solvers specifically designed for the sparse recovery
scenarios (e.g. \cite{vandenberg2007pr},\cite{figueiredo2007gps},
\cite{daubechies2008irw}).

\subsection{Decoding by $\ell^p$ minimization}
We have so far seen that with appropriate encoders, the decoders
$\Delta_1^\epsilon$ provide robust and stable recovery for
compressible signals even when the measurements are noisy
\cite{CRT05}, and that $(A_\omega,\Delta_1)$ is (2,2) instance optimal
in probability \cite{Wojtaszczyk08} when $A_\omega$ is an appropriate
random matrix. In particular, stability and robustness properties are conditioned on
an appropriate RIP while the instance optimality property is dependent
on the draw of the encoder matrix (which is typically called the {\em
  measurement matrix}) from an appropriate distribution, in addition
to RIP.

Recall that the decoders $\Delta_1$ and $\Delta_1^\epsilon$ were
devised because their action can be computed by solving convex
approximations to the combinatorial optimization problem
\eqref{eq:problem_l0} that is required to compute $\Delta_0$. The
decoders defined by \begin{align}
%P_1:  \min_{x} \|{x}\|_1 \text{ subject to }
%{b}={Ax}, 
\Delta_p^\epsilon({b})&:=\arg\min_y\|{y}\|_p \text{ s.t. }
{\|Ay-b\|_2\leq\epsilon}, \ \text{and} \label{eq:decoder_lp_noisy}
\\
\Delta_p(b)&:=\arg\min_y\|{y}\|_p \text{ s.t. } {Ay=b}, \label{eq:decoder_lp}
\end{align}
with $0<p<1$ are also approximations of $\Delta_0$, the actions of which
are computed via non-convex optimization problems that can be
solved, at least locally, still much faster than
\eqref{eq:problem_l0}.
It is natural to ask whether the decoders $\Delta_p$ and
$\Delta^{\epsilon}_p$ possess robustness, stability, and instance
optimality properties similar to those of $\Delta_1$ and
$\Delta^{\epsilon}_1$, and whether these are obtained under weaker
conditions on the measurement matrices than the analogous ones with
$p=1$.

Early work by Gribonval and co-authors
\cite{gribonval07:_highl,gribonval04:highlysparsICA04,gribonval06:simpletest,gribonval05:simpletesticassp}
take some initial steps in answering these questions. In particular,
they devise metrics that lead to sufficient conditions for uniqueness
of $\Delta_1({b})$ to imply uniqueness of $\Delta_p({b})$ and
specifically for having $\Delta_p({b})=\Delta_1({b})=x$. The authors
also present stability conditions in terms of various norms that bound
the error, and they conclude that the smaller the value of $p$ is, the
more non-zero entries can be recovered by \eqref{eq:decoder_lp}.
These conditions, however, are hard to check explicitly and no class
of deterministic or random matrices was shown to satisfy them at least
with high probability. On the other hand, the authors provide lower
bounds for their metrics in terms of generalized mutual
coherence. Still, these conditions are pessimistic in the sense that
they generally guarantee recovery of only very sparse vectors.

Recently, Chartrand showed that in the noise-free setting, a
sufficiently sparse signal can be recovered perfectly with $\Delta_p$,
where $0<p<1$, under less restrictive RIP requirements than
those needed to guarantee perfect recovery with $\Delta_1$. The
following theorem was proved in \cite{chartrand07letters}.  

{\thm {\em \cite{chartrand07letters}} \label{thm:P_rick} Let $0<p\le
 1$, and let $S\in \N$. Suppose that ${x}$ is
 $S$-sparse, and set ${b}={A}{x}$. If $A$ satisfies $\delta_{kS}
 +k^{\frac{2}{p}-1}\delta_{(k+1)S} < k^{\frac{2}{p}-1}-1$ for some
 $k>1$ such that $k\in \frac{1}{S} \N$, then $\Delta_p({b})=x$.}

Note that, for example, when $p=0.5$ and $k=3$, the above theorem only
requires $\delta_{3S} +27\delta_{4S} < 26$ to guarantee perfect
recovery with $\Delta_{0.5}$, a less restrictive condition than the
analogous one needed to guarantee perfect reconstruction with
$\Delta_{1}$, i.e., $\delta_{3S} +3\delta_{4S} < 2.$ Moreover, in
\cite{chartrand2008rip}, Staneva and Chartrand study a modified RIP
that is defined by replacing $\|Ac\|_2$ in \eqref{eq:UUP} with
$\|Ac\|_p$. They show that under this new definition of $\delta_S$,
the same sufficient condition as in Theorem \ref{thm:P_rick}
guarantees perfect recovery. Steneva and Chartrand also show that if
$A$ is an $M \times N$ Gaussian matrix, their sufficient condition is
satisfied provided $M > C_1(p)S + pC_2(p)S \log(N/S)$, where $C_1(p)$
and $C_2(p)$ are given explicitly in
\cite{chartrand2008rip}. 
It is important to note is that $pC_2(p)$ goes to zero as $p$ goes to
zero.  In other words, the dependence on $N$ of the required number of
measurements $M$ (that guarantees perfect recovery for all $x \in
\Sigma^N_S$) disappears as $p$ approaches 0. This result motivates a
more detailed study to understand the properties of the decoders
$\Delta_p$ in terms of stability and robustness, which is the
objective of this paper.

\subsubsection{Algorithmic Issues}
Clearly, recovery by $\ell^p$ minimization poses a non-convex
optimization problem with many local minimizers.  It is encouraging
that simulation results from recent papers, e.g.,
\cite{chartrand07letters,saab2008ssa}, strongly indicate that simple
modifications to known approaches like iterated reweighted least
squares algorithms and projected gradient algorithms yield $x^*$ that
are the global minimizers of the associated $\ell^p$ minimization
problem (or approximate the global optimizers very well). It is also
encouraging to note that even though the results presented in this
work and in others
\cite{chartrand07letters,gribonval07:_highl,gribonval04:highlysparsICA04,gribonval06:simpletest,gribonval05:simpletesticassp,saab2008ssa}
assume that the global minimizer has been found, a significant set of
these results, including all results in this paper, continue to hold
if we could obtain a feasible point $\widetilde{x}^*$ which satisfies
$\|\widetilde{x}^*\|_p \leq \|x\|_p$ (where $x$ is the vector to be
recovered). Nevertheless, it should be stated that to our knowledge,
the modified algorithms mentioned above have only been shown to
converge to local minima.

\subsection{Paper Outline}

In what follows, we present generalizations of the above results,
giving stability and robustness guarantees for $\ell^p$ minimization.
In Section \ref{sec:robust_stable} we show that the decoders
$\Delta_p$ and $\Delta_p^\epsilon$ are robust to noise and (2,p)
instance optimal in the case of appropriate measurement matrices. For
this section we rely and expand on our note \cite{saab2008ssa}. In
Section \ref{sec:instance_optimal} we extend \cite{Wojtaszczyk08} and
show that for the same range of dimensions as for decoding by $\ell^1$
minimization, i.e., when $A_\omega \in \R^{M\times N}$ with $M >
cS\log(N)$, $(A_\omega,\Delta_p)$ is also (2,2) instance optimal in
probability for $0<p<1$, provided the measurement matrix $A_\omega$ is
drawn from an appropriate distribution. The generalization follows the
proof of Wojtaszczyk in \cite{Wojtaszczyk08}; however it is
non-trivial and requires a variant of a result by Gordon and Kalton
\cite{Kalton} on the Banach-Mazur distance between a $p$-convex body
and its convex hull. In Section \ref{sec:numerical_results} we present
some numerical results, further illustrating the possible benefits of
using $\ell^p$ minimization and highlighting the behavior of the
$\Delta_p$ decoder in terms of stability and robustness. Finally, in
Section \ref{sec:proofs} we present the proofs of the main theorems
and corollaries.

While writing this paper, we became aware of the work of Foucart and
Lai \cite{Foucart08} which also shows similar $(2,p)$ instance
optimality results for $0<p<1$ under different sufficient
conditions. In essence, one could use the $(2,p)$-results of Foucart
and Lai to obtain $(2,2)$ instance optimality in probability results
similar to the ones we present in this paper, albeit with different
constants. Since neither the sufficient conditions for $(2,p)$
instance optimality presented in \cite{Foucart08} nor the ones in this
paper are uniformly weaker, and since neither provide uniformly better
constants, we simply use our estimates throughout.

\section{Main Results}

In this section, we present our theoretical results on the ability
of $\ell^p$ minimization to recover sparse and compressible signals in
the presence of noise. 
\subsection{Sparse recovery with $\Delta_p$: stability and
 robustness}\label{sec:robust_stable}
We begin with a deterministic stability and robustness theorem for
decoders $\Delta_p$ and $\Delta_p^\epsilon$ when $0<p<1$ that
generalizes Theorem \ref{thm:P2} of Cand{\`e}s et al. Note the
associated sufficient conditions on the measurement matrix, given in
\eqref{eq:theorem_cond} below, are weaker for smaller values of $p$
than those that correspond to $p=1$. The results in this subsection
were initially reported, in part, in \cite{saab2008ssa}.

In what follows, we say that a matrix $A$ satisfies the property
$P(k,S,p)$ if it satisfies
\begin{equation}
\delta_{kS} +k^{\frac{2}{p}-1}\delta_{(k+1)S} < k^{\frac{2}{p}-1} - 1,
\label{eq:theorem_cond}
\end{equation}
for $S\in \N$ and $k>1$ such that $k\in \frac{1}{S} \N$.
\begin{thm}[General Case] \label{thm:Pp_general}Let $0<p\le
  1$. Suppose that ${x}$ is arbitrary and $b=Ax+e$ where $\|e\|_2
  \leq \epsilon$. If $A$ satisfies $P(k,S,p)$, then
\begin{equation} \|\Delta_p^\epsilon({b}) - {x}\|_2^{p} \leq C_1
\epsilon^p +C_2\frac{\sigma_S(x)_{\ell^p}^p}{S^{1-p/2}}, 
\label{eq:error_bound}\end{equation}
where
\begin{equation}
C_1 =  2^p  \frac{{1+{k^{p/2-1}(2/p-1)^{-p/2}}}}{(1-\delta_{(k+1)S})^{p/2}-(1+\delta_{kS})^{p/2}k^{p/2-1}} ,\quad \text{and }
\label{eq:C1}\end{equation}
\begin{equation}
C_2 =  \frac{2(\frac{p}{2-p})^{p/2} }{k^{1-p/2}}\left( 1+ \frac{((2/p-1)^\ph+k^{p/2-1})(1+\delta_{kS})^{p/2}}{(1-\delta_{(k+1)S})^{p/2}-\frac{(1+\delta_{kS})^{p/2}}{k^{1-p/2}}} \right).
\label{eq:C2}\end{equation}
\end{thm}

\begin{rem}{\normalfont By setting $p=1$ and
$k=3$ in Theorem \ref{thm:Pp_general}, we obtain Theorem \ref{thm:P2},
with precisely the same constants.}\end{rem}

\begin{rem} {\normalfont The constants in Theorem \ref{thm:Pp_general}
    are generally well behaved; e.g., $C_1=5.31$ and $C_2=4.31$ for
    $\delta_{4S}=0.5$ and $p=0.5$. Note for $\delta_{4S}=0.5$ the
    sufficient condition \eqref{eq:theorem_cond} is not satisfied when
    $p=1$, and thus Theorem \ref{thm:Pp_general} does not yield any
    upper bounds on $\|\Delta_1(b)-x\|_2$ in terms of $\sigma_S(x)_{\ell^1}$.}
\end{rem}

{\cor[($2,p)$ instance optimality] \label{corr_io}Let $0<p\le 1$. Suppose that $A$ satisfies $P(k,S,p)$. Then $(A,\Delta_p)$ is
 $(2,p)$ instance optimal of order $S$ with constant $C_2^{1/p}$ where
 $C_2$ is as in \eqref{eq:C2}.}

{\cor[sparse case] \label{thm:Pp} Let $0<p\le 1$. Suppose $x\in
 \Sigma^N_S$ and ${b}={Ax}+{e}$ where $\|{e}\|_2 \leq \epsilon$. If $A$
 satisfies $P(k,S,p)$, then
\begin{equation} \|\Delta_p^\epsilon({b}) - {x}\|_2 \leq \left(C_1\right)^{1/p} \epsilon \nonumber,\end{equation}
where $C_1$ is as in \eqref{eq:C1}.}

\begin{rem} {\normalfont Corollaries \ref{corr_io} and \ref{thm:Pp}
    follow from Theorem \ref{thm:Pp_general} by setting $\epsilon=0$
    and $\sigma_S(x)_{\ell^p}=0$, respectively. Furthermore, Corollary
    \ref{thm:Pp} can be proved independently of Theorem
    \ref{thm:Pp_general} leading to smaller constants. See
    \cite{saab2008ssa} for the explicit values of these improved
    constants. Finally, note that setting $\epsilon=0$ in Corollary
    \ref{thm:Pp}, we obtain Theorem \ref{thm:P_rick} as a corollary.}
\end{rem}

\begin{rem} {\normalfont
In \cite{Foucart08}, Foucart and Lai give different sufficient conditions for exact recovery than those we present. In particular, they show that if
\begin{equation}
\delta_{mS}< {g}(m):=\frac{4(\sqrt{2}-1)(m/2)^{1/p-1/2}}{4(\sqrt{2}-1)(m/2)^{1/p-1/2}+2} \label{eq:FL_cond}
\end{equation}
holds for some $m \geq 2, m\in\frac{1}{S}\N$, then $\Delta_p$ will recover signals in $\Sigma_S^N$ exactly. 
Note that the sufficient condition in this paper, i.e., \eqref{eq:theorem_cond}, holds when
\begin{equation}
\delta_{mS}< {f}(m):=\frac{(m-1)^{2/p-1}-1}{(m-1)^{2/p-1}+1} \label{eq:SY_cond}
\end{equation}
for some $m \geq 2, m \in \frac{1}{S}\N$. In Figure \ref{fig:SY_vs_FL}, we
compare these different sufficient conditions as a function of $m$ for
$p=0.1,0.5,$ and $0.9$ respectively.  Figure~\ref{fig:SY_vs_FL}
indicates that neither sufficient condition is weaker than the other
for all values of $m$. In fact, we can deduce that \eqref{eq:FL_cond}
is weaker when $m$ is close to 2, while \eqref{eq:SY_cond} is weaker
when $m$ starts to grow larger. Since both conditions are only
sufficient, if either one of them holds for an appropriate $m$, then
$\Delta_p$ recovers all signals in $\Sigma_S^N$.
\begin{figure}[!ht]
\begin{center}
  \includegraphics[width=12cm]{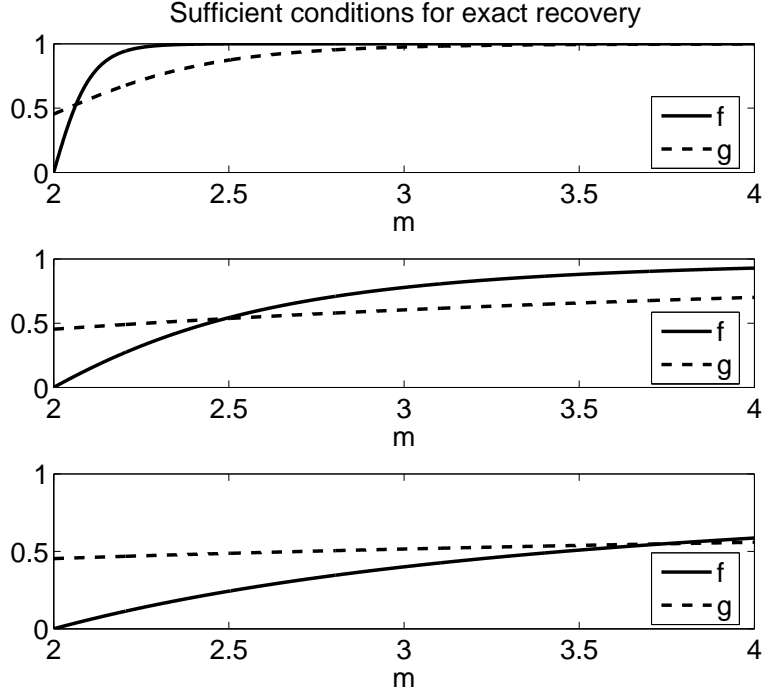}  
\end{center}
\caption{A comparison of the sufficient conditions on $\delta_{mS}$ in \eqref{eq:SY_cond} and \eqref{eq:FL_cond} as a function of $m$, for $p=0.1$ (top), $p=0.5$ (center) and $p=0.9$ (bottom).	\label{fig:SY_vs_FL}}
\end{figure}}
\end{rem}
\begin{rem}{\normalfont In \cite{davies:ric}, Davies and Gribonval
    showed that if one chooses $\delta_{2S} > \delta(p)$ (where
    $\delta(p)$ can be computed implicitly for $0<p\le 1$), then there
    exist matrices (matrices in $\R^{(N-1)\times N}$ that correspond
    to tight Parseval frames in $\R^{N-1}$) with the prescribed
    $\delta_{2S}$ for which $\Delta_p$ fails to recover signals in
    $\Sigma_S^N$. Note that this result does not contradict with the
    results that we present in this paper: we provide sufficient
    conditions (e.g., \eqref{eq:theorem_cond}) in terms of
    $\delta_{(k+1)S}$, where $k>1$ and $kS \in \N$, that
    \emph{guarantee} recovery by $\Delta_p$. These conditions are
    weaker than the corresponding conditions ensuring recovery by
    $\Delta_1$, which suggests that using $\Delta_p$ can be
    beneficial. Moreover, the numerical examples we provide in Section
    \ref{sec:numerical_results} indicate that by using $\Delta_p$,
    $0<p<1$, one can indeed recover signals in $\Sigma_{S}^N$, even when
    $\Delta_1$ fails to recover them (see Figure
    \ref{fig:phase_diagrams}).} 
\end{rem}

\begin{rem} {\normalfont\label{rem:sp} In summary, Theorem \ref{thm:Pp_general}
 states that if \eqref{eq:theorem_cond} is satisfied then we can
 recover signals in $\Sigma_S^N$ stably by decoding with
 $\Delta_p^\epsilon$.  It is worth mentioning that the sufficient
 conditions presented here reduce the gap between the conditions for
 exact recovery with $\Delta_0$ (i.e., $\delta_{2S}<1$) and with
 $\Delta_1$, e.g., $\delta_{3S}<1/3$. For example for $k=2$ and
 $p=0.5$, $\delta_{3S}<7/9$ is sufficient. In the next subsection, we
 quantify this improvement.}
\end{rem}

\subsection{The relationship between $S_1$ and $S_p$}
Let $A$ be an $M \times N$ matrix and suppose $\delta_m$, $m
\in\{1,\dots,\lfloor M/2 \rfloor\}$ are its $m$-restricted isometry
constants. Define $S_p$ for $A$ with $0<p\le 1$ as the largest value
of $S \in \N$ for which the slightly stronger version of
\eqref{eq:theorem_cond} given by
\begin{equation}
  \delta_{(k+1)S} <\frac{k^{\frac{2}{p}-1} - 1}{k^{\frac{2}{p}-1} + 1}\label{eq:cond_strong_p}
\end{equation}
holds for some $k>1$, $k \in \frac{1}{S} \N$. Consequently, by Theorem
\ref{thm:Pp_general}, $\Delta_p(Ax) = x$ for all $x\in
\Sigma^N_{S_p}$. We now establish a relationship between $S_1$ and
$S_p$.
\begin{prop}\label{prop:rel_Sp_S1} 
  Suppose, in the above described setting, there exists $S_1 \in \N$
  and $k>1$, $k\in \frac{1}{S_1}\N$ such that
\begin{equation}
\delta_{(k+1)S_1} <\frac{k - 1}{k + 1}\label{eq:cond_strong}
\end{equation}
Then $\Delta_1$ recovers all $S_1$-sparse vectors, and $\Delta_p$
recovers all $S_p$ sparse vectors with 
$$
S_p = \left \lfloor \frac{k+1}{k^{\frac{p}{2-p}}+1} S_1 \right \rfloor.
$$ 
\end{prop}
\begin{rem}\normalfont
For example, if $\delta_{5S_1}<3/5$ then using $\Delta_{\frac{2}{3}}$, we can recover all $S_{\frac{2}{3}}$-sparse vectors with $S_{\frac{2}{3}}=\lfloor \frac{5}{3}S_1 \rfloor$.
\end{rem}
\subsection{Instance optimality in probability and  $\Delta_p$}\label{sec:instance_optimal}
In this section, we show that $(A_\omega,\Delta_p)$ is $(2,2)$
instance optimal in probability when $A_\omega$ is an appropriate
random matrix. Our approach is based on that of \cite{Wojtaszczyk08},
which we summarize now.  A matrix $A$ is said to possess the
$\LQ_1(\alpha)$ property if and only if
$${A}(B_1^N) \supset \alpha B_2^M,$$%
where $B_q^n$ denotes the $\ell^q$ unit ball in $\R^n$. In
\cite{Wojtaszczyk08}, Wojtaszczyk shows that random Gaussian matrices
of size $M\times N$ as well as matrices whose columns are drawn
uniformly from the sphere possess, with high probability, the
$\LQ_1(\alpha)$ property with $\alpha = \mu
\sqrt{\frac{\log{(N/M)}}{M}}$. Noting that such matrices also satisfy
$\RIP((k+1)S,\delta)$ with $S<c\frac{M}{log{(N/M)}}$, again with high
probability, Wojtaszczyk proves that $\Delta_1$, for these matrices,
is (2,2) instance optimal in probability of order $S$. Our strategy
for generalizing this result to $\Delta_p$ with $0<p<1$ relies on
a generalization of the $\LQ_1$ property to an $\LQ_p$
property. Specifically, we say that a matrix $A$ satisfies
$\LQ_p(\alpha)$ if and only if
$$ {A}(B_p^N) \supset \alpha B_2^M. $$
We first show that a random matrix $A_\omega$, either Gaussian or
uniform as mentioned above, satisfies the $\LQ_p(\alpha)$ property with
$$\alpha
=\frac{1}{C(p)}\left(\mu^2
{\frac{\log{(N/M)}}{M}}\right)^{(1/p-1/2)}.
$$ 
Once we establish this property, the proof of instance optimality in
probability for $\Delta_p$ proceeds largely unchanged from
Wojtaszczyk's proof with modifications to account only for the
non-convexity of the $\ell^p$-quasinorm with $0<p<1$. 

Next, we present our results on instance optimality of the $\Delta_p$
decoder, while deferring the proofs to Section \ref{sec:proofs}.
Throughout the rest of the paper, we focus on two classes of random
matrices: ${A}_\omega$ denotes $M\times N$ matrices, the entries of
which are drawn from a zero mean, normalized column-variance Gaussian
distribution, i.e., ${A}_\omega=(a_{i,j})$ where $a_{i,j}\sim
\mathcal{N}(0,1/\sqrt{M})$; in this case, we say that $A_\omega$ is an
$M\times N$ Gaussian random matrix. ${\widetilde{A}}_\omega$, on the
other hand, denotes $M\times N$ matrices, the columns of which are
drawn uniformly from the sphere; in this case we say that
$\widetilde{A}_\omega$ is an $M\times N$ uniform random matrix. In
each case, $(\Omega,P)$ denotes the associated probability space.

We start with a lemma (which generalizes an analogous result of
\cite{Wojtaszczyk08}) that shows that the matrices $A_\omega$ and
$\widetilde{A}_\omega$ satisfy the $\LQ_p$ property with high
probability.

\begin{lemma} \label{lemma:LQ_p} Let $0<p\le 1$, and let $A_\omega$
 be an $M\times N$ Gaussian random matrix. For $0<\mu<1/\sqrt{2}$,
 suppose that $K_1 M (\log M)^\xi \leq N \leq e^{K_2 M}$ for some
 $\xi>(1-2\mu^2)^{-1}$ and some constants $K_1,K_2 > 0$. Then, there
 exists a constant $c=c(\mu,\xi,K_1,K_2)>0$, independent of $p$, $M$,
 and $N$, and a set
$$\Omega_\mu=\left\{\omega\in \Omega: {A}_\omega(B_p^N)\supset \frac{1}{C(p)}\left(\mu^2{\frac{\log{N/M}}{M}}\right)^{1/p-1/2}B_2^M\right\}\newline$$
such that $P(\Omega_\mu)\geq 1-e^{-cM}$.

In other words, $A_\omega$ satisfies the
${\normalfont \LQ}_p(\alpha)$, $\alpha =1/C(p)
\left(\mu^2\frac{{\log{(N/M)}}}{M}\right)^{1/p-1/2}$, with probability
$\geq 1-e^{-cM}$ on the draw of the matrix.  Here $C(p)$ is a positive
constant that depends only on $p$. (In particular, $C(1)=1$ and see
\eqref{eq:C_p} for the explicit value of $C(p)$ when $0<p<1$).
This statement is true also for ${\widetilde{A}}_\omega$.
\end{lemma}

The above lemma for $p=1$ can be found in \cite{Wojtaszczyk08}.  As we
will see in Section~\ref{sec:proofs}, the generalization of this
result to $0<p<1$ is non-trivial and requires a result from
\cite{Kalton}, cf. \cite{Litvak00}, relating certain ``distances'' of
$p$-convex bodies to their convex hulls. It is important to note that
this lemma provides the machinery needed to prove the following
theorem, which extends to $\Delta_p$, $0<p<1$, the analogous result of
Wojtaszczyk \cite{Wojtaszczyk08} for $\Delta_1$.

In what follows, for a set $T\subseteq \{1,\dots,N\}$, $T^c:=
\{1,\dots,N\}\setminus T$; for $y\in\R^N$, $y_T$ denotes the vector
with entries $y_T(j)=y(j)$ for all $j\in T$, and $y_T(j)=0$ for $j \in
T^c$.

\begin{thm} \label{thm:inst_opt1} 
  Let $0<p<1$. Suppose that ${A}\in\R^{M\times N}$ satisfies
  ${\normalfont\RIP}(S,\delta)$ and
  ${\normalfont \LQ}_p\left(\frac{1}{C(p)}(\mu^2/S)^{1/p-1/2}\right)$ 
  for some $\mu >0$ and $C(p)$ as in \eqref{eq:C_p}. Let $\Delta$ be
  an arbitrary decoder. If $({A},\Delta)$ is (2,p) instance optimal of
  order $S$ with constant $C_{2,p}$, then for any ${x} \in \R^N$ and
  ${e} \in \R^M$, all of the following hold.
\begin{enumerate}
\item $\|\Delta({Ax+e}) -{x}  \|_2 \leq C(\| {e}\|_2 +
\frac{\sigma_S(x)_{\ell^p}}{S^{1/p-1/2}})$ \smallskip
\item $\|\Delta({Ax}) -{x}  \|_2  \leq C(\| {Ax}_{T_0^c}\|_2 + \sigma_S(x)_{\ell^2})$\smallskip
\item $\|\Delta({Ax+e}) -{x}  \|_2 \leq C(\| {e}\|_2 + \sigma_S(x)_{\ell^2}+\|{Ax}_{T_0^c}\|_2)$
\end{enumerate}
Above, $T_0$ denotes the set of indices of the largest (in magnitude)
$S$ coefficients of $x$; the constants (all denoted by $C$) depend on
$\delta$, %$\beta(?)$,
$\mu$, $p$, and $C_{2,p}$ but not on $M$ and $N$. For the explicit
values of these constants see \eqref{eq:IO_consts1} and
\eqref{eq:IO_consts2}.
\end{thm}
Finally, our main theorem on the instance optimality in probability of
the $\Delta_p$ decoder follows. 

\begin{thm} Let $0<p<1$, and let $A_\omega$ be an $M\times N$ Gaussian
 random matrix. Suppose that $N \ge M [\log(M)]^2$. There exists constants
 $c_1,c_2,c_3>0$ such that for all $S\in \N$ with $S \leq c_1
 M/\log{(N/M)}$, the following are true.
\begin{enumerate}
\item There exists $\Omega_1$ with $P(\Omega_1) \geq 1-3e^{-c_2 M}$ such that
for all $\omega \in \Omega_1$
\begin{equation}
\|\Delta_p(A_\omega(x)+e)-{x}\|_2 \leq C(\|{e}\|_2 + \frac{\sigma_S(x)_{\ell^p}}{S^{1/p-1/2}} ),
\label{eq:i}
\end{equation}
for any $x\in \R^N$ and for any $e\in \R^M$. 
\item For any ${x} \in \R^N$, there exists $\Omega_x$ with $P(\Omega_x) \geq
1-4e^{-c_3 M}$ such that for all $\omega \in \Omega_x$
\begin{equation}
\|\Delta_p(A_\omega(x)+e)-{x}\|_2 \leq C\left(\|{e}\|_2 + \sigma_S(x)_{\ell^2} \right),
\label{eq:ii}
\end{equation}
for any $e\in \R^M$.
\end{enumerate}
The statement also holds for $\widetilde{A}_\omega$, i.e., for random
matrices the columns of which are drawn independently from a uniform
distribution on the sphere.\label{thm:inst_opt_final}
\end{thm}

\begin{rem}{\normalfont The constants above (both denoted by $C$)
    depend on the parameters of the particular $\LQ_p$ and $\RIP$
    properties that the matrix satisfies, and are given explicitly in
    Section \ref{sec:proofs}, see \eqref{eq:IO_consts1} and
    \eqref{eq:IOconsts_2}. The constants $c_1,c_2$, and $c_3$ depend
    only on $p$ and the distribution of the underlying random matrix
    (see the proof in Section \ref{thm_pf}) and are independent of $M$
    and $N$.}
\end{rem}

\begin{rem} {\normalfont
 Clearly, the statements do not make sense if the hypothesis of the
 theorem forces $S$ to be 0. In turn, for a given $(M,N)$ pair, it is
 possible that there is no positive integer $S$ for which the
 conclusions of Theorem \ref{thm:inst_opt_final} hold. In particular,
 to get a non-trivial statement, one needs
 $M>\frac{1}{c_1}\log(N/M)$.}
\end{rem}
\begin{rem}{\normalfont
Note the difference in the order of the quantifiers between
conclusions (i) and (ii) of Theorem \ref{thm:inst_opt_final}.
Specifically, with statement (i), once the matrix is drawn from the
``good'' set $\Omega_1$, we obtain the error guarantee \eqref{eq:i}
for every $x$ and $e$. In other words, after the initial draw of a
good matrix $A$, stability and robustness in the sense of
\eqref{eq:i} are ensured. On the other hand, statement (ii)
concludes that associated with every $x$ is a ``good'' set
$\Omega_x$ (possibly different for different $x$) such that if the
matrix is drawn from $\Omega_x$, then stability and robustness in the
sense of \eqref{eq:ii} are guaranteed. Thus, in (ii), for every $x$,
a different matrix is drawn, and with high probability on that draw
\eqref{eq:ii} holds.}
\end{rem}

\begin{rem}{\normalfont The above theorem pertains to the decoders
    $\Delta_p$ which, like the analogous theorem for $\Delta_1$
    presented in \cite{Wojtaszczyk08}, requires no knowledge of the
    noise level. In other words, $\Delta_p$ provides estimates of
    sparse and compressible signals from limited and noisy
    observations without having to explicitly account for the noise in
    the decoding. This provides an improvement on Theorem
    \ref{thm:Pp_general} and a practical advantage when estimates of
    measurement noise levels are absent.}
\end{rem}

\section{Numerical Experiments}\label{sec:numerical_results}
In this section, we present some numerical experiments to highlight
important aspects of sparse reconstruction by decoding using
$\Delta_p$, $0<p\leq1$. First, we compare the sufficient conditions
under which decoding with $\Delta_p$ guarantees perfect recovery of
signals in $\Sigma^N_S$ for different values of $p$ and $S$. Next, we
present numerical results illustrating the robustness and instance
optimality of the $\Delta_p$ decoder. Here, we wish to observe the
linear growth of the $\ell^2$ reconstruction error
$\|\Delta_p({Ax+e})-{x}\|_2$, as a function of $\sigma_S(x)_{\ell^2}$
and of $\|{e}\|_2$.

\begin{figure}[!ht]
\begin{center}
%\hspace*{3ex}
  \includegraphics[width=6cm,origin=c]{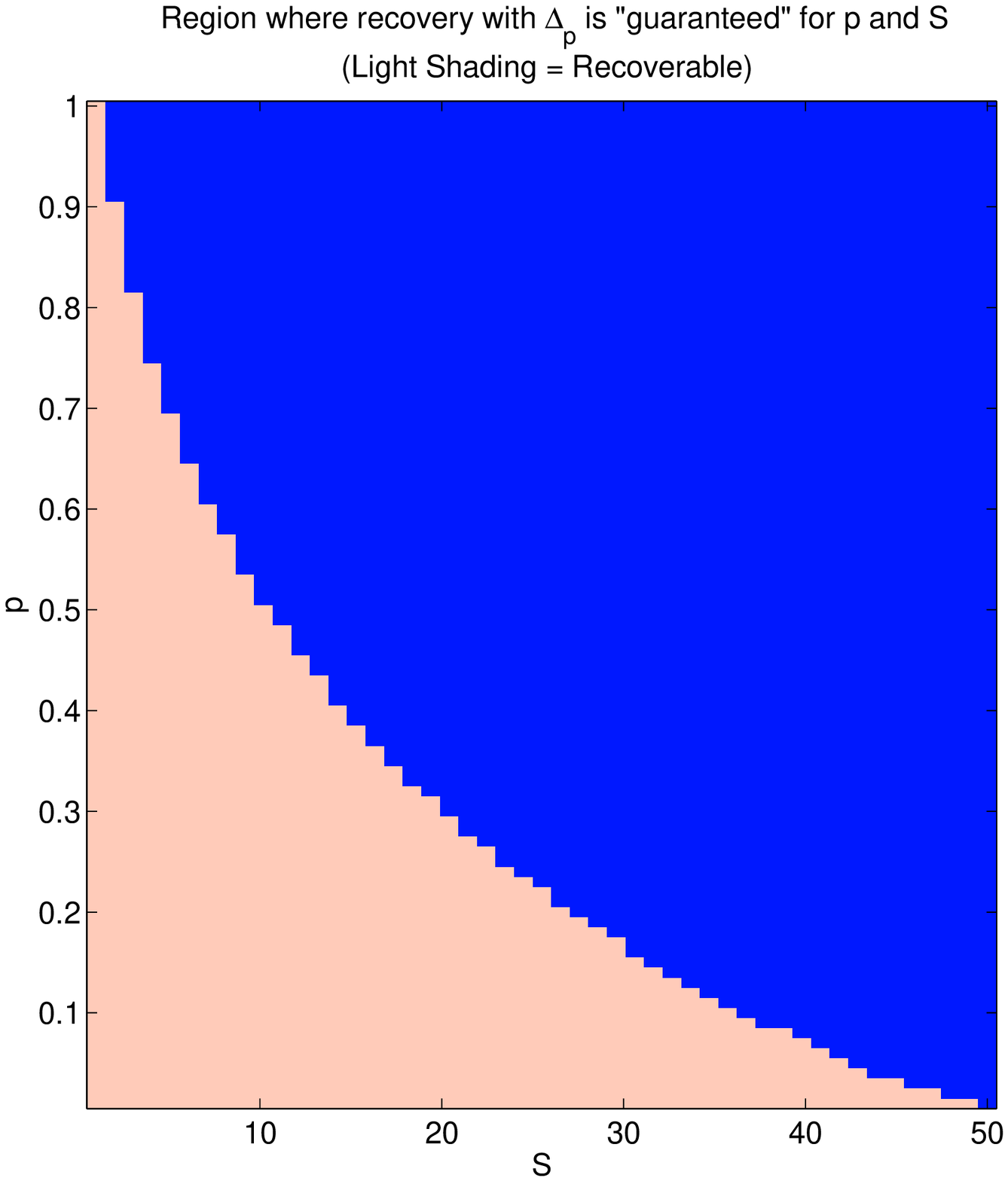}  
  \includegraphics[width=6cm,origin=c]{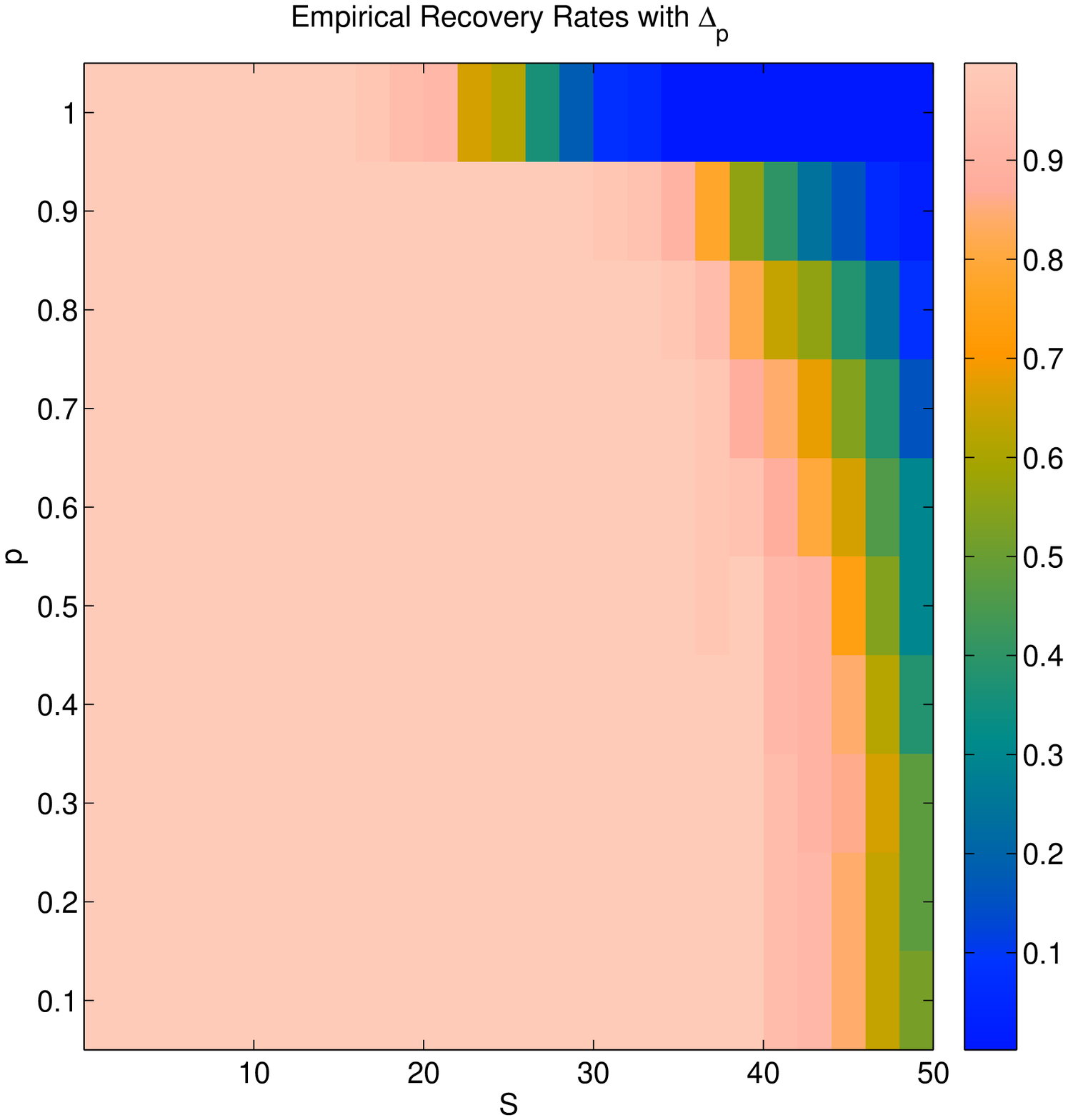}  
\end{center}
\caption{For a Gaussian matrix ${A}\in {{{{\mathbb{R}}}}}^{100 \times 300}$, whose $\delta_S$ values are estimated via MC simulations, we generate the theoretical (left) and practical (right) phase-diagrams for reconstruction via $\ell^p$ minimization. \label{fig:phase_diagrams}}
\end{figure}

To that end, we generate a $100\times 300$ matrix $A$ whose columns are
drawn from a Gaussian distribution and we estimate its RIP constants
$\delta_S$ via Monte Carlo (MC) simulations. Under the assumption that
the estimated constants are the correct ones (while in fact
they are only lower bounds), Figure \ref{fig:phase_diagrams} (left) shows the
regions where \eqref{eq:theorem_cond} guarantees recovery for
different $(S,p)$-pairs. %Moreover, given the value of $k$ which
%guarantees the recovery via $\Delta_1$ of the largest number of
%non-zeros $S_1$, we plot (yellow line) the estimate of $S_p$ from
%Corollary \ref{cor:cor1}. The green line shows the most optimistic
%value of $S_p$ possible, i.e., [CAN'T RECALL OUR RATIONAL HERE, OZGUR?].
On the other hand, Figure \ref{fig:phase_diagrams} (right) shows the
empirical recovery rates via $\ell^p$ quasinorm minimization: To
obtain this figure, for every $S=1,\dots,49$, we chose 50 different
instances of $x\in \Sigma^{300}_S$ where non-zero coefficients of each
were drawn i.i.d. from the standard Gaussian distribution. These
vectors were encoded using the same measurement matrix $A$ as
above. Since there is no known algorithm that will yield the global
minimizer of the optimization problem \eqref{eq:decoder_lp}, we
approximated the action of $\Delta_p$ by using a projected gradient
algorithm on a sequence of smoothed versions of the $\ell^p$
minimization problem: In \eqref{eq:decoder_lp}, instead of minimizing
the $\|y\|_p$, we minimized
$\left(\sum_i{({y}_i^2+\epsilon^2)^{p/2}}\right)^{1/p}$ initially with
a large $\epsilon$. We then used the corresponding solution as the
starting point of the next subproblem obtained by decreasing the value
of $\epsilon$ according to the rule $\epsilon_n=(0.99)\epsilon_{n-1}$. We continued reducing the value of $\epsilon$ and
solving the corresponding subproblem until $\epsilon$ becomes very
small.
Note that this approach is similar to the one described in
\cite{chartrand07letters}.  The empirical results show that $\Delta_p$
(in fact, the approximation of $\Delta_p$ as described above) is
successful in a wider range of scenarios than those predicted by
Theorem \ref{thm:Pp_general}. This can be attributed to the fact that
the conditions presented in this paper are only sufficient, or to the
fact that in practice what is observed is not necessarily a
manifestation of uniform recovery. Rather, the practical results could
be interpreted as success of $\Delta_p$ with high probability on
either $x$ or ${A}$.

\begin{figure}[!htbp]
\begin{center}
  \includegraphics[width=6cm,origin=c]{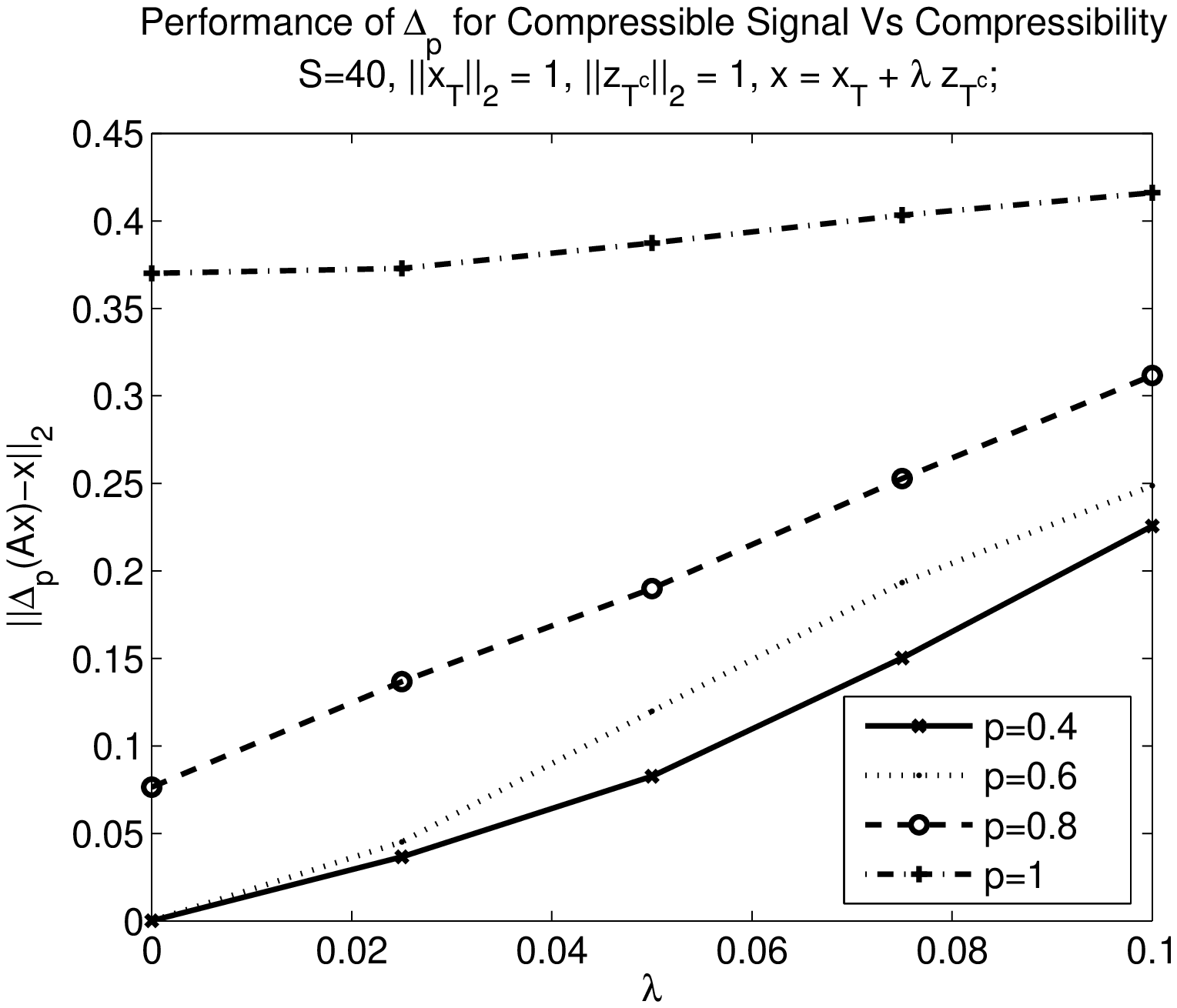}  
  \includegraphics[width=6cm,origin=c]{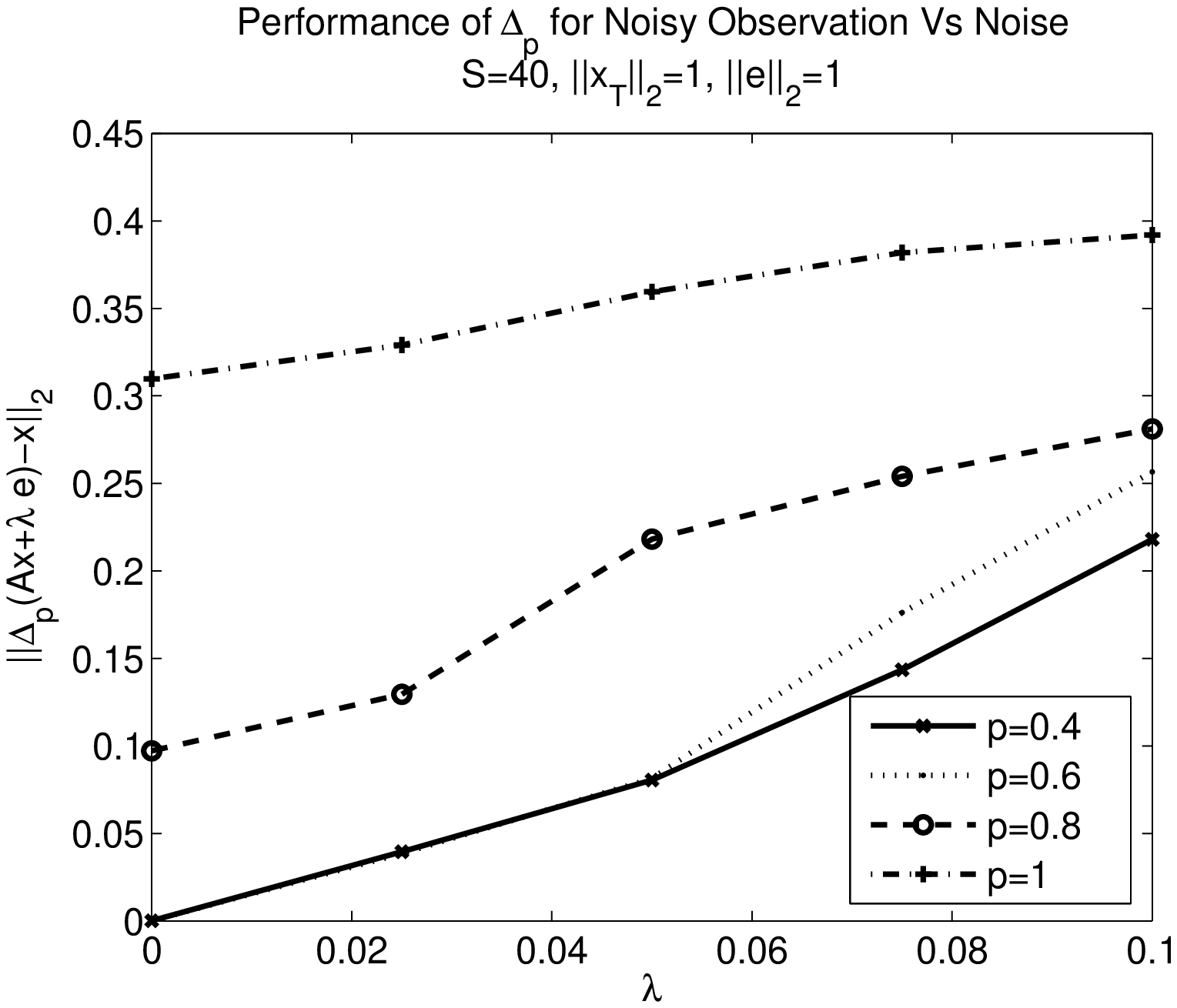}  
\end{center}
\caption{Reconstruction error with compressible signals (left), noisy observations (right). Observe the almost linear growth of the error in compressible signals and for different values of $p$, highlighting the instance optimality of the decoders. The plots were generated by averaging the results of 10 experiments with the same matrix $A$ and randomized locations of the coefficients of $x$. 
\label{fig:inst_opt}
}
\end{figure}

Next, we generate scenarios that allude to the conclusions of Theorem
\ref{thm:inst_opt_final}. To that end, we generate a signal composed
of $x_{T} \in \Sigma_{40}^{300}$, supported on an index set $T$, and a
signal $z_{T^c}$ supported on $T^c$, where all the coefficients are
drawn from the standard Gaussian distribution. We then normalize
$x_{T}$ and $z_{T^c}$ so that $\|x_T\|_2=\|z_{T^c}\|_2=1$ and generate
$x = x_T + \lambda z_{T^c}$ with increasing values of $\lambda$
(starting from 0), thereby increasing
$\sigma_{40}(x)_{\ell^2}\approx \lambda$. For this experiment, we choose our
measurement matrix $A \in \R^{100\times300}$ by drawing its columns
uniformly from the sphere. For each value of $\lambda$ we measure the
reconstruction error $\|\Delta_p(Ax)-x\|_2$, and we repeat the process
10 times while randomizing the index set $T$ but preserving the
coefficient values. We report the averaged results in Figure
\ref{fig:inst_opt} (left) for different values of $p$. Similarly, we
generate noisy observations $Ax_T + \lambda e$, of a sparse signal
$x_T \in \Sigma_{40}^{300}$ where $\|x_T\|_2=\|e\|_2 = 1$ and we increase
the noise level starting from $\lambda=0$. Here, again, the
non-zero entries of $x_T$ and all entries of $e$ were chosen i.i.d.
from the standard Gaussian distribution and then the vectors were
properly normalized. Next, we measure $\|\Delta_p(Ax_T+\lambda
e)-x_T\|_2$ (for 10 realizations where we randomize $T$) and report
the averaged results in Figure \ref{fig:inst_opt} (right) for
different values of $p$. In both these experiments, we observe that
the error increases roughly linearly as we increase $\lambda$,
i.e., $\sigma_{40}(x)_{\ell^2}$ and the noise power,
respectively. Moreover, when the signal is highly compressible or when
the noise level is low, we observe that reconstruction using $\Delta_p$ with
$0<p<1$ yields a lower approximation error than that with $p=1$. It is also worth noting 
that for values of $p$ close to one, even in the case of sparse signals with no noise, the
average reconstruction error is non-zero. This may be due to the fact that for such large $p$
the number of measurements is not sufficient for the recovery of signals with $S=40$, further highlighting the benefits of using the 
decoder $\Delta_p$, with smaller values of $p$.

\begin{figure}[!htbp]
\begin{center}
  \includegraphics[width=6cm,height=5cm,origin=c]{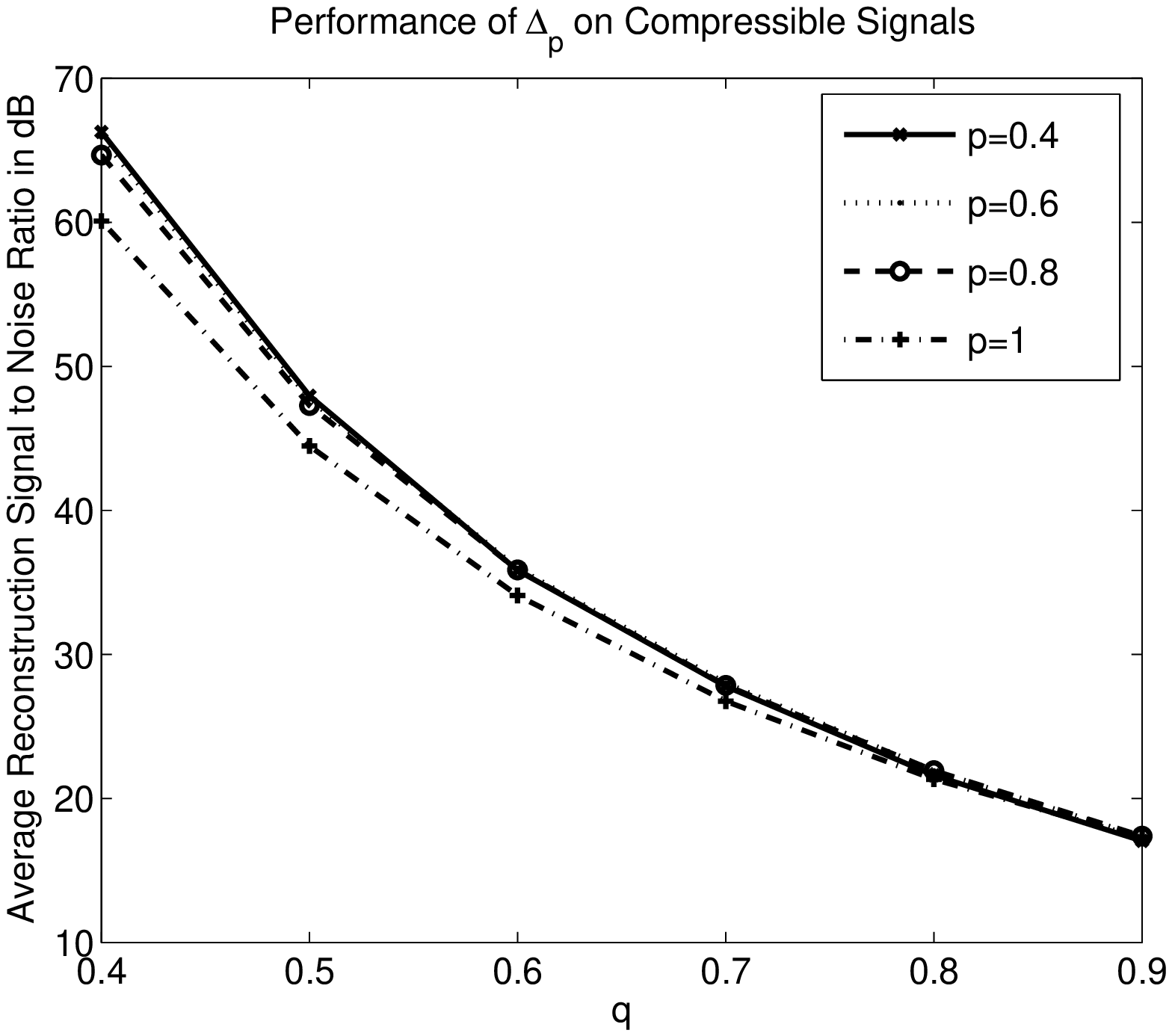}  
  \includegraphics[width=6cm,height=5cm,origin=c]{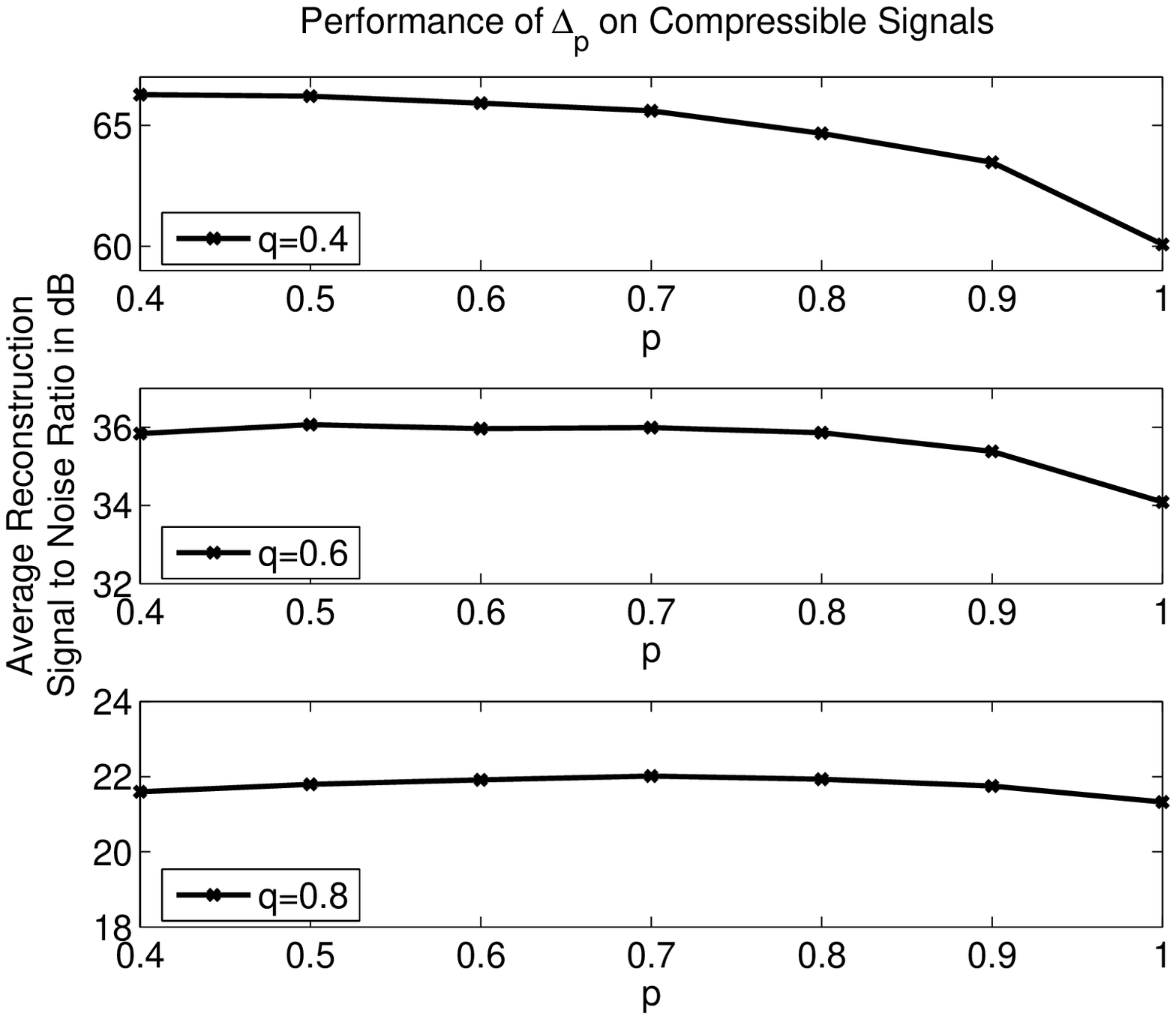}
\end{center}
\caption{Reconstruction signal to noise ratios (in dB) obtained by
  using $\Delta_p$ to recover signals whose sorted coefficients decay
  according to a power law ($x(j)=c j^{-1/q}, \|x\|_2=1$) as a
  function of $q$ (left) and as a function of $p$ (right). The
  presented results are averages of 50 experiments performed with
  different matrices in $\R^{100\times200}$. Observe that for highly
  compressible signals, e.g., for $q=0.4$, there is a $5$~dB gain in
  using $p<0.6$ as compared to $p=1$. The performance advantage is
  about $2$~dB for $q=0.6$. As the signals become much less
  compressible, i.e., as we increase $q$ to $0.9$ the performances are
  almost identical.\label{fig:p_q} }%
\end{figure}

Finally, in Figure \ref{fig:p_q}, we plot the results of an experiment
in which we generate signals $x\in \R^{200}$ with sorted coefficients
$x(j)$ that decay according to some power law. In particular, for
various values of $0<q<1$, we set $x(j) =c j^{-1/q}$ such that
$\|x\|_2=1$. We then encode $x$ with 50 different $100 \times 200$
measurement matrices the columns of which were drawn from the uniform
distribution on the sphere, and examine the approximations obtained by
decoding with $\Delta_p$ for different values of $0<p<1$. The results
indicate that values of $p \approx q$ provide the lowest
reconstruction errors. Note that in Figure \ref{fig:p_q}, we report
the results in form of signal to noise ratios defined
as $$SNR=20\log_{10}\left(\frac{\|x\|_2}{\|\Delta{(Ax)}-x\|_2}\right).$$

\section{Proofs}\label{sec:proofs}

\subsection{\bf Proof of Proposition \ref{prop:rel_Sp_S1}} 

First, note that for any $A\in \R^{M\times N}$, $\delta_m$ is
non-decreasing in $m$. Also, the map $k \mapsto \frac{k-1}{k+1}$ is
increasing in $k$ for $k \ge 0$. 

Set 
$$L:=(k+1)S_1, \quad \lt=k^{\frac{p}{2-p}}, \quad \text{and
$\Spt=\frac{L}{\lt+1}$.}
$$
Then
$$
\delta_{(\lt+1)\Spt}= \delta_{(k+1)S_1} <
\frac{k-1}{k+1}=\frac{\lt^\frac{2-p}{p}-1}{\lt^\frac{2-p}{p}+1}.
$$
We now describe how to choose $\ell$ and $S_p$ such that $\ell
\ge \lt$, $S_p\in \N$, and \mbox{$(\ell+1)S_p = L$} (this will be
sufficient to complete the proof using the monotonicity observations
above). First, note that this last equality is satisfied only if
$(\ell,S_p)$ is in the set
$$\{(\frac{n}{L-n},L-n):\ n=1,\dots,L-1\}.
$$
Let $n^*$ be such that 
\begin{equation}\label{nstar}
\frac{n^*-1}{L-n^*+1} < \lt \le \frac{n^*}{L-n^*}.
\end{equation}
To see that such an $n^*$ exists, recall that $\lt=k^{\frac{p}{2-p}}$
where $0<p<1$. Also, $(k+1)S_1=L$ with $S_1\in \N$, and
$k>1$. Consequently, $1< \lt < k \le L-1$, and $k \in
\{\frac{n}{L-n}:\ n=\lceil \frac{L}{2}\rceil,\dots, L-1\}$. Thus, we
know that we can find $n^*$ as above. Furthermore,
$\frac{n^*}{L-n^*}>1$. It follows from \eqref{nstar} that
$$L-n^* \le \Spt < L-n^*+1. 
$$
We now choose 
$$\ell=\frac{n^*}{L-n^*}, \quad \text{and}\ S_p=\lfloor \Spt \rfloor =
L-n^*.
$$
Then $(\ell+1)S_p = L$, and $\ell \ge \lt$. So, we conclude that for
$\ell$ as above and   
$$
S_p=\lfloor \Spt \rfloor = \left \lfloor
\frac{k+1}{k^{\frac{p}{2-p}}+1}S_1 \right \rfloor,
$$
we have 
$$ \delta_{(\ell+1)S_p} < \frac{\ell^\frac{2-p}{p}-1}{\ell^\frac{2-p}{p}+1}.
$$
Consequently, the condition of Corollary \ref{thm:Pp} is satisfied and
we have the desired conclusion.
\qed

%\noindent{\bf Proof outline for Theorem \ref{thm:Pp}.}
\subsection{\bf Proof of Theorem \ref{thm:Pp_general}} 
We modify the proof of Cand{\`e}s et. al. of the analogous result for
the encoder $\Delta_1$ (Theorem 2 in \cite{CRT05}) to account for the
non-convexity of the $\ell^p$ quasinorm. We give the full proof for
completeness. We stick to the notation of \cite{CRT05} whenever
possible.

Let $0<p<1$, $x\in \R^N$ be arbitrary, and define
$x^*:=\Delta_p^\epsilon(b)$ and $h:=x^*-x$. Our goal is to
obtain an upper bound on $\|h\|_2$ given that $\|Ah\|_2\le 2\epsilon$ (by
definition of $\Delta_p^\epsilon$).

Below, for a set $T\subseteq \{1,\dots,N\}$, $T^c:=
\{1,\dots,N\}\setminus T$; for $y\in\R^N$, $y_T$ denotes the vector
with entries $y_T(j)=y(j)$ for all $j\in T$, and $y_T(j)=0$ for $j \in
T^c$.

\noindent
(\hskip 0.5mm I\hskip 0.5mm) \hskip 0.5mm We start by decomposing $h$
as a sum of sparse vectors with disjoint support. In particular,
denote by $T_0$ the set of indices of the largest (in magnitude) $S$
coefficients of $x$ (here $S$ is to be determined later). Next,
partition $T_o^c$ into sets $T_1,T_2,\dots$, $|T_j|=L$ for $j\ge 1$
where $L\in \N$ (also to be determined later), such that $T_1$ is the
set of indices of the $L$ largest (in magnitude) coefficients of
$h_{T_0^c}$, $T_2$ is the set of indices of the second $L$ largest
coefficients of $h_{T_0^c}$, and so on. Finally let $T_{01}:=T_0 \cup
T_1$. We now obtain a lower bound for $\|A h \|_2^p$ using the RIP constants
of the matrix $A$. In particular, we have
\begin{eqnarray}
\|Ah\|_2^p &=& \|Ah_{T_{01}} + \sum_{j\ge 2} Ah_{T_j} \|_2^p \nnl
&\ge& \| Ah_{T_{01}}\|_2^p - \sum_{j\ge 2} \|Ah_{T_j} \|_2^p \nnl
&\ge& (1-\delta_{L+|T_0|})^{p/2} \|h_{T_{01}}\|_2^p -
(1+\delta_{L})^{p/2} \sum_{j\ge 2} \|h_{T_j}\|_2^p. \label{eqAh}
\end{eqnarray}
Above, together with RIP, we used the fact that $\|\cdot\|_2^p$
satisfies the triangle inequality for any $0<p<1$.  What now remains
is to relate $\|h_{T_{01}}\|_2^p$ and $\sum_{j\ge 2} \|h_{T_j}\|_2^p$
to $\|h\|_2$.

\noindent
(\hskip 0.5mm II\hskip 0.5mm ) \hskip 0.5mm Next, we aim to bound $\sum_{j\ge 2} \|h_{T_j}\|_2^p$ from above in terms
of $\|h\|_2$. To that end, we proceed as in \cite{CRT05}. First, note
that $|h_{T_{j+1}}(\ell)|^p \le |h_{T_j}(\ell')|^p$ for all $\ell \in T_{j+1}, \ell' \in T_j$, and
thus $|h_{T_{j+1}}(\ell)|^p \le \|h_{T_j}\|_p^p/L$. It follows that
$\|h_{T_{j+1}}\|_2^2 \le L^{1-\frac{2}{p}} \|h_{T_j}\|_p^2$, and
consequently
\begin{equation}\label{eqsum1}
\sum_{j\ge 2} \|h_{T_j}\|_2^p \le L^{\ph-1}\sum_{j\ge 1}
\|h_{T_j}\|_p^p=L^{\ph-1} \|h_{T_o^c}\|_p^p.
\end{equation}
Next, note that, similar to the case when $p=1$ as shown in
\cite{CRT05}, the ``error'' $h$ is concentrated on the ``essential
support'' of $x$ (in our case $T_0$). To quantify this claim, we
repeat the analogous calculation in \cite{CRT05}: Note, first, that by
definition of $x^*$, 
$$
\|x^*\|_p^p =
\|x+h\|_p^p=\|x_{T_0}+h_{T_0}\|_p^p+\|x_{T^c_0}+h_{T^c_0}\|_p^p \le \|x\|_p^p.
$$
As $\|\cdot\|_p^p$ satisfies the triangle inequality, we then have
$$
\|x_{T_0}\|_p^p-\|h_{T_0}\|_p^p+\|h_{T_0^c}\|_p^p-\|x_{T_0^c}\|_p^p \le \|x\|_p^p.
$$
Consequently,
\begin{equation}\label{crt_pf_1}
\|h_{T_o^c}\|_p^p \le \|h_{T_0}\|_p^p + 2 \|x_{T_0^c}\|_p^p,
\end{equation}
which, together with \eqref{eqsum1}, implies
\begin{equation}
\sum_{j\ge 2} \|h_{T_j}\|_2^p \le L^{\ph-1}(\|h_{T_{0}}\|_p^p+
2\|x_{T_0^c}\|_p^p) 
\le \rho^{1-\ph}(\|h_{T_{01}}\|_2^p+ 2
|T_0|^{\ph-1}\|x_{T_0^c}\|_p^p), \label{eqsum2}
\end{equation}
where $\rho:=\frac{|T_0|}{L}$, and we used the fact that
$\|h_{T_0}\|_p^p \le |T_0|^{1-\ph} \|h_{T_0}\|_2^p$ (which follows as
$|\text{supp}(h_{T_0})|=|T_0|$). Using \eqref{eqsum2} and
\eqref{eqAh}, we obtain 
\begin{equation}
\|Ah\|_2^p \ge C_{p,L,|T_0|} \|h_{T_{01}}\|_2^p - 2 \rho^{1-\ph}
|T_0|^{\ph-1} (1+\delta_L)^{\ph} \|x_{T_0^c}\|_p^p,
\end{equation}
where 
\begin{equation}
C_{p,L,|T_0|}:=(1-\delta_{L+|T_0|})^\ph-(1+\delta_L)^\ph \rho^{1-\ph}.
\end{equation}
At this point, using $\|Ah\|_2 \le 2\epsilon$, we obtain an upper bound
on $\|h_{T_{01}}\|_2$ given by
\begin{equation} \label{upbd1}
\|h_{T_{01}}\|_2^p \le \frac{1}{C_{p,L,|T_0|}}\left ( (2\epsilon)^p + 2 \rho^{1-\ph}
(1+\delta_L)^{\ph}\frac{\|x_{T_0^c}\|_p^p}{|T_0|^{1-\ph}} \right),
\end{equation}
provided $C_{p,L,|T_0|}>0$ (this will impose the condition given in
\eqref{eq:theorem_cond} on the RIP constants of the underlying matrix
$A$).

\noindent
(\hskip 0.5mm III\hskip 0.5mm ) \hskip 0.5mm To complete the proof, we will show that the error vector $h$
is concentrated on $T_{01}$. 
Denote by $h_{T_0^c}[m]$ the $m$th largest (in
magnitude) coefficient of $h_{T_0^c}$ and observe that $|h_{T_0^c}[m]|^p
\le \|h_{T_0^c}\|_p^p/m$. As $h_{T_{01}^c}[m]=h_{T_0^c}[L+m]$, we then
have
\begin{equation} \label{crt_pf_2}
\|h_{T_{01}^c}\|_2^2=\sum_{m\ge L+1} |h_{T_0^c}[m]|^2 \le \sum_{m\ge
L+1} \left(\frac{\|h_{T_0^c}\|^p_p}{m}\right)^\frac{2}{p}\le \frac{\|h_{T_0^c}\|_p^2}{L^{\hp-1}(2/p-1)}.
\end{equation}
Here, the last inequality follows because for $0<p<1$
$$\sum_{m\ge L+1} m^{-\hp} \le \int_L^\infty t^{-\hp} dt =
\frac{1}{L^{\hp-1}(2/p-1)}.
$$
Finally, we use \eqref{crt_pf_1} and \eqref{crt_pf_2} to conclude
\begin{eqnarray}
\|h\|_2^2&=&\|h_{T_{01}}\|_2^2+\|h_{T_{01}^c}\|_2^2 
\le \|h_{T_{01}}\|_2^2 + \left[ \frac{\|h_{T_0}\|_p^p +
2\|x_{T^c_{0}}\|_p^p}{L^{1-\ph}(2/p-1)^\ph}\right]^{\hp} \nnl
&\le& \left[ \big(1+\rho^{1-\ph}(2/p-1)^{-\ph}\big)\|h_{T_{01}}\|_2^p +2\rho^{1-\ph}(2/p-1)^{-\ph}\frac{\|x_{T^c_{0}}\|_p^p}{|T_0|^{1-\ph}}\right]^{\hp}. \label{upbd2}
\end{eqnarray}
Above, we used the fact that $\|h_{T_0}\|_p^p \le
|T_0|^{1-\ph}\|h_{T_0}\|_2^p$, and that for any $a,b\ge 0$, and 
$\alpha\ge 1$, $a^\alpha+b^\alpha \le (a+b)^\alpha$.

\noindent
(\hskip 0.5mm IV\hskip 0.5mm ) \hskip 0.5mm  We now set $|T_0|=S$, $L=kS$ where $k$ and $S$ are chosen such
that $C_{p,kS,S}>0$ which is equivalent to having $k$, $S$, and $p$ satisfy 
\eqref{eq:theorem_cond}. In this case,
$\|x_{T_0^c}\|_p=\sigma_S(x)_{\ell^p}$, $\rho=1/k$, and  
combining
\eqref{upbd1} and \eqref{upbd2} yields 
\begin{equation}
\|h\|_2^p \le C_1 \epsilon^p + C_2\frac{\sigma_S(x)_{\ell^p}^p}{S^{1-\ph}}
\end{equation}
where $C_1$ and $C_2$ are as in \eqref{eq:C1} and \eqref{eq:C2},
respectively. 
\qed

\subsection{\bf Proof of Lemma \ref{lemma:LQ_p}.}
(\hskip 0.5mm I\hskip 0.5mm) The following result of Wojtaszczyk
\cite[Proposition 2.2]{Wojtaszczyk08} will be
useful.
\begin{prop}[\cite{Wojtaszczyk08}]\label{prop:Woj} 
 Let $A_\omega$ be an $M\times N$ Gaussian random matrix, let
 $0<\mu<1/\sqrt{2}$, and suppose that $K_1 M (\log M)^\xi \leq N
 \leq e^{CM}$ for some $\xi>(1-2\mu^2)^{-1}$ and some constants
 $K_1,K_2 > 0$. Then, there exists a constant
 $c=c(\mu,\xi,K_1,K_2)>0$, independent of $M$ and $N$, and a set
$$\Omega_\mu=\left\{\omega: {A}_\omega(B_1^N)\supset \mu\sqrt{\frac{\log{N/M}}{M}}B_2^M\right\}$$
such that  
$$P(\Omega_\mu)\geq 1-e^{-cM}.$$
The above statement is true also for ${\widetilde{A}}_\omega$. 
\end{prop}

We will also use the following adaptation of \cite[Lemma 2]{Kalton}
for which we will first introduce some notation. Define a {\em body}
to be a compact set containing the origin as an interior point and
star shaped with respect to the origin \cite{Litvak00}. Below, we use
$conv(K)$ to denote the convex-hull of a body
$K$. 
For $K\subseteq B$, we denote by $d_1(K,B)$ the ``distance'' between
$K$ and $B$ given by
$$d_1(K,B):=\inf\{\lambda>0 :\ K\subset B
\subset \lambda K\} = \inf\{\lambda>0 :\ \frac{1}{\lambda}B\subset K
\subset B\}. $$ Finally, we call a body $K$ {\em $p$-convex} if for
any $x,y \in K$, $\lambda x + \mu y \in K$ whenever $\lambda, \mu \in
[0,1]$ such that $\lambda^p +\mu^p = 1$.
{\lemma{Let $0<p<1$, and let $K$
be a $p$-convex body in $\R^n$. If $conv(K) \subset B_2^n$, then
%$$d_1(K,B) \leq   (c/p)^{2/p^2-2/p} d_1(conv(K),B_2^n)^{(2/p-1)},$$
$$d_1(K,B_2^n) \leq   C(p) d_1(conv(K),B_2^n)^{(2/p-1)},$$ where %$$C_p=\frac{((2-p)/p)^{(2-p)/p^2}}{((1-p) \ln{(2)})^{(2-2p)/p^2}}.$$
$$C(p)=\left(2^{1-p}+\frac{(1-p)2^{1-p/2}}{p}\right)^\frac{2-p}{p^2}\left(\frac{1}{(1-p)\log 2}\right)^\frac{2-2p}{p^2} .$$
\label{lem:Litvak}}}
We defer the proof of this lemma to the Appendix.
%\vskip 3mm

\noindent
(\hskip 0.5mm II\hskip 0.5mm ) Note that ${\widetilde{A}}_\omega(B_1^N)\subset B_2^M$. This
follows because $\|{\widetilde{A}}_\omega\|_{1\rightarrow 2}$, which
is equal to the largest column norm of $\widetilde{A}_\omega$, is 1
by construction. Thus, for $x\in B_1^N$,
$$
\|\widetilde{A}_\omega(x)\|_2 \le
\|{\widetilde{A}}_\omega\|_{1\rightarrow 2}\|x\|_1 \le 1,
$$
that is, ${\widetilde{A}}_\omega(B_1^N)\subset B_2^M$, and so
$d_1(\widetilde{A}_\omega(B_1^N),B^M_2)$ is well-defined. Next, by
Proposition \ref{prop:Woj}, we know that there exists $\Omega_\mu$ with
$P(\Omega_\mu)\ge 1-e^{-cM}$ such that for all $\omega \in \Omega_\mu$,
\begin{equation}\label{incl1}
{\widetilde{A}}_\omega(B_1^N)\supset \mu\sqrt{\frac{\log{N/M}}{M}}B_2^M
\end{equation}
From this point on, let $\omega\in \Omega_\mu$. Then
$$B_2^M \supset{\widetilde{A}}_\omega(B_1^N)\supset\mu\sqrt{\frac{\log{N/M}}{M}}B_2^M,$$
and consequently
\begin{equation}
d_1({\widetilde{A}}_\omega(B_1^N),B_2^M) \leq \left(\mu\sqrt{\frac{\log{N/M}}{M}}\right)^{-1}.\label{eq:BM_dist_l1}
\end{equation}

The next step is to note that $conv(B_p^N) = B_1^N$ and consequently
$$conv\left({\widetilde{A}}_\omega(B_p^N)\right)=
{\widetilde{A}}_\omega \left(conv(B_p^N)\right)
=
{\widetilde{A}}_\omega(B_1^N).$$ 
We can now invoke Lemma \ref{lem:Litvak} to conclude that
\begin{align}
d_1({\widetilde{A}}_\omega(B_p^N),B_2^M) &\leq C(p)
d_1(conv({\widetilde{A}}_\omega(B_p^N)),B_2^M)^{\frac{2-p}{p}}\nonumber\\
&= C(p) d_1({\widetilde{A}}_\omega(B_1^N),B_2^M)^{\frac{2-p}{p}}.
\end{align}
Finally, by using (\ref{eq:BM_dist_l1}), we find that 
\begin{equation}
d_1({\widetilde{A}}_\omega(B_p^N),B_2^M) \leq C(p) \left(\mu^2{\frac{\log{N/M}}{M}}\right)^{1/2-1/p},
\end{equation}
and consequently
\begin{equation} {\widetilde{A}}_\omega(B_p^N)\supset \frac{1}{C(p)}
\left(\mu^2{\frac{\log{N/M}}{M}}\right)^{(1/p-1/2)}B_2^M.
\end{equation}
In other words, the matrix $\widetilde{A}_\omega$ has the
$\LQ_p(\alpha)$ property with % $\alpha =\frac{1}{C_p} \left(\mu^2
% \frac{{\log{(N/M)}}}{M}\right)^{1/p-1/2}$
the desired value of $\alpha$ for every $\omega \in \Omega_\mu$ with
$P(\Omega_\mu)\ge 1-e^{-cM}$. Here $c$ is as specified in Proposition
\ref{prop:Woj}.

To see that the same is true for ${A}_\omega$, note that there exists
a set $\Omega_0$ with $P(\Omega_0) >1 -e^{-cM}$ such that for all
$\omega \in \Omega_0$, $\|A_j(\omega)\|_2<2$, for every column ${A}_j$
of $A_\omega$ (this follows from RIP). Using this observation one can
trace the above proof with minor modifications.  \qed

\subsection{\bf Proof of Theorem \ref{thm:inst_opt1}.}
We start with the following lemma, the proof of which for $p<1$ follows
with very little modification from the analogous proof of Lemma 3.1 in
\cite{Wojtaszczyk08} and shall be omitted.

{\lemma \label{lmWoj} Let $0<p<1$ and suppose that ${A}$ satisfies $\RIP(S,\delta)$
and ${\normalfont \LQ}_p\left(\gamma_p/S^{1/p-1/2}\right)$ with
$\gamma_p:=\mu^{2/p-1}/C(p)$. Then for every ${x} \in \R^N$, there exists
${\widetilde{x}} \in \R^N$ such that
\renewcommand{\labelenumi}{(\roman{enumi})}
$$
{Ax} = {A\widetilde{x}}, \quad
\|{\widetilde{x}}\|_p \leq \frac{S^{1/p-1/2}}{\gamma_p}
\|{Ax}\|_2, \quad \text{and} \quad 
\|{\widetilde{x}}\|_2 \leq C_3\|{Ax}\|_2. 
$$
Here, $C_3= \frac{1}{\gamma_p}
+\frac{\gamma_p(1-\delta)+1}{(1-\delta^2)\gamma_p}$. Note that $C_3$
depends only on $\mu$, $\delta$ and $p$.}

We now proceed to prove Theorem \ref{thm:inst_opt1}. Our proof follows the steps of \cite{Wojtaszczyk08} and differs in the handling of the non-convexity of the $\ell^p$ quasinorms for $0<p<1$.

First, recall that $A$ satisfies $\RIP(S,\delta)$ and
$\LQ_p(\gamma_p/S^{1/p-1/2})$, so by Lemma \ref{lmWoj}, there exists
$z \in \R^N$ such that ${Az}=e$, $\|{z}\|_p \leq
\frac{S^{1/p-1/2}}{\gamma_p}\|{e}\|_2$, and $\|{z}\|_2 \leq
C_3\|{e}\|_2$. Now, ${A(x+z)}={Ax + e}$, and $\Delta$ is $(2,p)$ instance
optimal with constant $C_{2,p}$. Thus,
$$\|\Delta({A(x)+e})-{(x+z)} \|_2 \leq C_{2,p} \frac{\sigma_S(x+z)_{\ell^p}}{S^{1/p-1/2}},$$ and consequently
\begin{align}
\|\Delta({A(x)+e})-{x} \|_2 &\leq \|{z}\|_2 + C_{2,p} \frac{\sigma_S(x+z)_{\ell^p}}{S^{1/p-1/2}}\nonumber\\
&\leq C_3\|{e}\|_2 + C_{2,p} \frac{\sigma_S(x+z)_{\ell^p}}{S^{1/p-1/2}}\nonumber\\
&\leq C_3\|{e}\|_2 + 2^{1/p-1}C_{2,p} \frac{\sigma_S(x)_{\ell^p} +\|{z}\|_p}{S^{1/p-1/2}}\nonumber\\
&\leq C_3\|{e}\|_2 + 2^{1/p-1}C_{2,p} \frac{\sigma_S({x})_{\ell^p}
}{S^{1/p-1/2}}  +2^{1/p-1}C_{2,p}\frac{\|{e}\|_2}{\gamma_p}, \nonumber
\end{align}
where in the third inequality we used the fact in any that $\ell^p$
quasinorm satisfies the inequality $\|a+b\|_p \le
2^{\frac{1}{p}-1}(\|a\|_p+\|b\|_p)$ for all $a,b\in \R^N$. So, we
conclude
\begin{equation}
\|\Delta({A(x)+e})-{x} \|_2 \leq  \left(C_3+2^{1/p-1}C_{2,p}/\gamma_p\right)\|{e}\|_2 +2^{1/p-1}C_{2,p} \frac{\sigma_S({x})_{\ell^p} }{S^{1/p-1/2}}. \label{eq:IO_consts1}
\end{equation}
That is (i) holds with $C=C_3+2^{1/p-1}C_{2,p}(1/\gamma_p+1)$.

Next, we prove parts (ii) and (iii) of Theorem
\ref{thm:inst_opt1}. As in the analogous proof of \cite{Wojtaszczyk08}, Theorem
\ref{thm:inst_opt1} (ii) can be seen as a special case of Theorem
\ref{thm:inst_opt1} (iii), with ${e}=0$. We therefore turn to proving
(iii). 
%with the associated constant which will be specified below.
Once again, by Lemma \ref{lmWoj}, there exists $v$ and $z$ in $\R^N$
such that the following hold.
$$
\begin{array}{lll}
{Av=e}; &\|{v}\|_p \leq \frac{S^{1/p-1/2}}{\gamma_p}
\|{e}\|_2, & \|{v}\|_2 \leq C_3\|{e}\|_2, \ \ \text{and} \\
{Az=Ax_{T_0^c}}; &\|{z}\|_p \leq \frac{S^{1/p-1/2}}{\gamma_p}
\|{Ax_{T_0^c}}\|_2, &\|{z}\|_2 \leq C_3\|{Ax_{T_0^c}}\|_2. 
\end{array}
$$
Here $T_0$ is the set of indices of the largest (in magnitude) $S$
coefficients of $x$, and $T_0^c$ and $x_{T_o^c}$ are as in the
proof of Theorem \ref{thm:Pp_general}.

Similar to the previous part we can see that ${A(x_{T_0}+z+v)} =
{Ax+e} $ and by the hypothesis of $(2,p)$ instance
optimality of $\Delta$, we have $$\|\Delta({Ax+e})-({x_{T_0}+z+v})\|_2 \leq C_{2,p}
\frac{\sigma_S({x_{T_0}+z+v})_{\ell^p}}{S^{1/p-1/2}}.$$ Consequently
observing that ${x_{T_0}}={x-x_{T_0^c}}$ and using the triangle
inequality, 
\begin{align}
\|\Delta({A(x)+e})-{x} \|_2 &\leq \|{x_{T_0^c}-z-v}\|_2 + C_{2,p} \frac{\sigma_S({x_{T_0}+z+v})_{\ell^p}}{S^{1/p-1/2}}\nonumber\\
&\leq\|{x_{T_0^c}-z-v}\|_2 + 2^{1/p-1}(C_{2,p})\left( \frac{\|{z}\|_p+\|{v}\|_p}{S^{1/p-1/2}}\right)\nonumber\\
&\leq {\sigma_S({x})}_{\ell^2} +\|{z}\|_2+\|{v}\|_2 + 2^{1/p-1}C_{2,p}\left( {\frac{\|{Ax_{T_0^c}}\|_2}{\gamma_p}+\frac{\|{e}\|_2}{\gamma_p}}\right)\nonumber\\
&\leq {\sigma_S({x})}_{\ell^2} +\left(C_3+2^{1/p-1}\frac{C_{2,p}}{\gamma_p}\right)(\|{e}\|_2+\|{Ax_{T_0^c}}\|_2). 
\label{eq:IO_consts2}
\end{align}
That is (iii) holds with $C=1+C_3+2^{1/p-1}\frac{C_{2,p}}{\gamma_p}$. By setting $e=0$, one can see that this is the same constant associated with (ii).
This concludes the proof of this theorem. 
\qed

\subsection{\bf Proof of Theorem \ref{thm:inst_opt_final}.} \label{thm_pf}
%
%First, we show that $(A_\omega,\Delta_p)$ is $(2,p)$ instance optimal
%of order $S$ for an appropriate range of $S$ with high probability. To
%this end, by Theorem \ref{thm:Pp_general}, it is sufficient to show
%that the relevant RIP constants of $A_\omega$ satisfy
%\eqref{eq:theorem_cond} with high probability. This, in turn, follows
%from one of the fundamental results in compressed sensing theory: for
%any $\delta\in (0,1)$, there exists
%$\widetilde{c}_1,\widetilde{c}_2>0$ and $\Omega_{\RIP}$ with
%$P(\Omega_\text{RIP}) \ge 1-2e^{-\widetilde{c}_2M}$, all depending
%only on $\delta$, such that $A_\omega$, $\omega\in \Omega_\RIP$,
%satisfies $\RIP(\ell,\delta)$ for any $\ell \leq \widetilde{c}_1
%\frac{M}{log(N/M)}$. See, e.g.,
%\cite{CandesTao05},\cite{baraniuk100spr}, for the proof of this
%statement as well as for the explicit values of the constants. To
%simplify the remainder of the argument, we use
%\eqref{eq:cond_strong_p}, which implies \eqref{eq:theorem_cond}, and
%in \eqref{eq:cond_strong_p}, we set $k=2$. Now, choose $\delta \in
%(0,1)$ such that $\delta<\frac{2^{2/p-1}-1}{2^{2/p-1}+1}$. Then, with
%$\widetilde{c_1},\widetilde{c_2}$, and $\Omega_\RIP$ as above, for
%every $\omega \in \Omega_\RIP$ and for every
%$S<\frac{\widetilde{c}_1}{3}\frac{M}{log(N/M)}$, the RIP
%constants of $A_\omega$ satisfy \eqref{eq:cond_strong_p}, and thus by Theorem \ref{thm:Pp_general}
%$(A_\omega,\Delta_p)$ is instance optimal of order $S$ with constant
%$C_2^{1/p}$ as in \eqref{eq:C2}. 
First, we show that $(A_\omega,\Delta_p)$ is $(2,p)$ instance optimal
of order $S$ for an appropriate range of $S$ with high probability. One of the fundamental results in compressed sensing theory states that for
any $\delta\in (0,1)$, there exists
$\widetilde{c}_1,\widetilde{c}_2>0$ and $\Omega_{\RIP}$ with
$P(\Omega_\text{RIP}) \ge 1-2e^{-\widetilde{c}_2M}$, all depending
only on $\delta$, such that $A_\omega$, $\omega\in \Omega_\RIP$,
satisfies $\RIP(\ell,\delta)$ for any $\ell \leq \widetilde{c}_1
\frac{M}{log(N/M)}$. See, e.g.,
\cite{CandesTao05},\cite{baraniuk100spr}, for the proof of this
statement as well as for the explicit values of the constants. Now, choose $\delta \in (0,1)$ such that $\delta<\frac{2^{2/p-1}-1}{2^{2/p-1}+1}$. Then, with
$\widetilde{c_1},\widetilde{c_2}$, and $\Omega_\RIP$ as above, for
every $\omega \in \Omega_\RIP$ and for every
$S<\frac{\widetilde{c}_1}{3}\frac{M}{log(N/M)}$, the RIP
constants of $A_\omega$ satisfy \eqref{eq:cond_strong_p} (and hence \eqref{eq:theorem_cond}), with $k=2$. Thus, by %Theorem \ref{thm:Pp_general}
Corollary \ref{corr_io} $(A_\omega,\Delta_p)$ is instance optimal of order $S$ with constant
$C_2^{1/p}$ as in \eqref{eq:C2}.

Now, set $S_1=c_1\frac{M}{log(N/M)}$ with $c_1 \leq \widetilde{c}_1/3$
such that $S_1\in \N$ (note that such a $c_1$ exists if $M$ and $N$
are sufficiently large). By the hypothesis of the theorem, $M$ and $N$
satisfy the hypothesis of the Lemma \ref{lemma:LQ_p} with $\xi=2$, $K_1=1$,
some $0<\mu<1/2$, and an appropriate $K_2$ (determined by
$\widetilde{c}_1$ above). Because
$$
\left(\mu^2 \frac{log(N/M)}{M}\right)^{1/p-1/2} = \left(\mu^2
 \frac{c_1}{S_1}\right)^{1/p-1/2}
$$
by Lemma \ref{lemma:LQ_p}, there exists $\Omega_\mu$,
$P(\Omega_\mu)\ge 1-e^{-cM}$ such that for every $\omega \in
\Omega_\mu$, $A_\omega$ satisfies
$\LQ_p\left(\frac{\gamma_p(\mu)}{{S_1}^{1/p-1/2}}\right)$ where
$\gamma_p(\mu):= \frac{c_1^{1/p-1/2}\mu^{2/p-1}}{C(p)}$.
Consequently, set $\Omega_1:=\Omega_\text{RIP} \cap \Omega_\mu$. Then,
$P(\Omega_1)\ge 1-2e^{-\widetilde{c}_2 M}-e^{-cM}\ge 1-3e^{-c_2 M}$,
for $c_2=\min\{\widetilde{c}_2,c\}$. Note that $c_2$ depends on $c$, which is now a
universal constant, and $\widetilde{c}_2$, which depends only on the
distribution of $A_\omega$ (and in particular its concentration of
measure properties, see \cite{baraniuk100spr}). Now, if $\omega \in
\Omega_1$, $A_\omega$ satisfies $\RIP(3S_1,\delta)$, thus
$\RIP(S_1,\delta)$, as well as
$\LQ_p\left(\frac{\gamma_p}{{S_1}^{1/p-1/2}}\right)$.
Therefore we can apply part (i) of Theorem
\ref{thm:inst_opt1} to get the first part of this theorem, i.e.,
\begin{equation}\|\Delta({A_\omega(x)+e})-{x} \|_2 \leq C\left(
   \|e\|_2 +\frac{\sigma_{S_1}(x)_{\ell^p}}{{S_1}^{1/p-1/2}}
 \right).\end{equation} 
Here $C$ is as in \eqref{eq:IO_consts1} with $C_{2,p}=C_2^{1/p}$.  To finish the
proof of part (i), note that for $S\leq S_1$,
$\sigma_{S_1}(x)_{\ell^p}\leq \sigma_{S}(x)_{\ell^p}$ and $S^{1/p-1/2}
\leq S_1^{1/p-1/2}$.

To prove part (ii), first define $T_0$ as the support of the $S_1$
largest coefficients (in magnitude) of $x$ and
$T_0^c=\{1,...,N\}\setminus {T_0}$.  Now, note that for any $x$ there
exists a set $\widetilde{\Omega}_x$ with $P(\widetilde{\Omega}_x) \geq
1 - e^{-\tilde{c}M}$ for some universal constant $\widetilde{c}>0$,
such that for all $\omega \in \widetilde{\Omega}_x$, $\|{A_\omega
 x_{T_0^c}}\|_2\leq 2\|{x_{T_0^c}}\|_2 = 2\sigma_{S_1}{(x)_{\ell^2}}$
(this follows from the concentration of measure property of Gaussian
matrices, see, e.g., \cite{baraniuk100spr}). Define
$\Omega_x:=\widetilde{\Omega}_x \cap \Omega_1$. Thus, $P(\Omega_x)\geq
1-3e^{-c_2 M}-e^{-\widetilde{c}M}\geq 1-4e^{-c_3 M}$ where
$c_3=\min\{c_2,\widetilde{c}\}$. Note that the dependencies of $c_3$
are identical to those of $c_2$ discussed above. Recall that for
$\omega \in \Omega_1$, $A_\omega$ satisfies both $\RIP(S_1,\delta)$
and $\LQ_p\left(\frac{\gamma_p}{{(S_1)}^{1/p-1/2}}\right)$. We can now
apply part (iii) of Theorem \ref{thm:inst_opt1} to obtain for $\omega
\in \Omega_x$
\begin{equation}\|\Delta({A_\omega(x)+e})-{x} \|_2 \leq C\left({3\sigma_{S_1}({x})}_{\ell^2} + \|{e}\|_2\right).
\label{eq:IOconsts_2}\end{equation}
Above, the constant $C$ is as in \eqref{eq:IO_consts2}. Once again, note that for $S\leq S_1$, $\sigma_{S_1}(x)_{\ell^2}\leq \sigma_{S}(x)_{\ell^2}$ to finish the proof for any $S\leq S_1$. 
\qed

\section{Appendix: Proof of Lemma \ref{lem:Litvak}}
In this section we provide the proof of Lemma \ref{lem:Litvak} for the
sake of completeness and also because we explicitly calculate the
optimal constants involved. Let us first introduce some notation used
in \cite{Kalton} and \cite{Litvak00}.

For a body $K \subset \R^n$, define its {\em gauge functional} by 
$\|x\|_K:= \inf\{t>0 : \ x \in tK\}$,
and let $T_q(K)$, $q \in (1,2]$, be the
smallest constant $C$ such that 
$$\forall m \in \N, \ x_1,...,x_m \in K \quad \inf_{\epsilon_i=\pm1}\left\{\|\sum_{i=1}^m\epsilon_i x_i\|_K\right\}\leq C m^{1/q}.$$ 
Given a $p$-convex body $K$ and a positive integer $r$, define
$$\alpha_r = \alpha_r(K) := \sup\{\frac{\|\sum_{i=1}^r x_i\|_K}{r}: \
x_i \in K, i\leq r\}.
$$ 
Note that $\alpha_r \le r^{-1+1/p}$.

Finally, conforming with the notation used in \cite{Kalton} and \cite{Litvak00}, we define $\delta_K := d_1(K,conv(K))$. Note that this should not cause confusion as we do not refer to the RIP constants throughout the rest of the paper. 
It can be shown  by a result of \cite{Peck81} that $\delta_K= \sup_r
\alpha_r(K)$, cf. \cite[Lemma 1]{Kalton} for a proof.
 
%\begin{rem} Henceforth,  to conform with the notation used in \cite{Kalton} and \cite{Litvak00}, we use $\delta_K$ to refer to $d_1(K,conv(K))$, and {\em not} to the restricted isometry constants. \end{rem}

We will need the following propositions. 

\begin{prop}[sub-additivity of $\|\cdot\|_K^p$] For the gauge
functional $\|\cdot\|_K$ associated with a $p$-convex body $K \in
\R^n$, the following inequality holds for any $x, y \in \R^n$. 
\begin{equation} \label{eq:subadd}
\|x+y \|_K^p \leq \|x\|_K^p + \|y\|_K^p.
\end{equation}
\end{prop}
\begin{pf}
Let $r=\|x\|_K$ and $u=\|y\|_K$. If at least one of $r$ and $u$ is
zero, then \eqref{eq:subadd} holds trivially. (Note that, as $K$ is
a body, $\|x\|_K=0$ if and only if $x=0$.) So, we may assume that
both $r$ and $u$ are strictly positive. Since $K$ is compact, it
follows that $x/r \in K$ and $y/u \in K$. Furthermore, $K$ is
$p$-convex, i.e., for all $\alpha, \beta \in [0,1]$ with $\alpha +
\beta =1$, we have $\alpha^{1/p} x/r + \beta^{1/p} y/u \in K$. In
particular, choose $\alpha = \frac{r^p}{r^p+u^p}$ and $\beta =
\frac{u^p}{r^p+u^p}$. This gives $\frac{x+y}{(r^p+u^p)^{1/p}} \in
K$. Consequently, by the definition of the gauge functional
$\|\frac{x+y}{(r^p+u^p)^{1/p}}\|_K\leq 1$. Finally,
$\|\frac{x+y}{(r^p+u^p)^{1/p}}\|_K^p =
\frac{\|{x+y}\|_K^p}{(r^p+u^p)}\leq 1$ and $\|x+y\|_K^p\leq
r^p+u^p=\|x\|_K^p+\|y\|_K^p$. \qed
\end{pf}
\begin{prop}\label{prop_fin} $T_2(B_2^n)=1$.
\end{prop}
\begin{pf}
Note that $\|\cdot\|_{B_2^n}=\|\cdot\|_2$, and thus, by definition,
$T_2(B_2^n)$ is the smallest constant $C$ such that for every
positive integer $m$ and for every choice of points $x_1,...,x_m \in
B_2$,
\begin{equation}
%\inf_{\epsilon_i=\pm1}\left{\|\sum_{i=1}^m \epsilon_i x_i\|_2\right} \leq C \sqrt{m}.
\inf_{\epsilon_i=\pm1}\left\{\|\sum_{i=1}^m\epsilon_i x_i\|_2\right\}\leq C \sqrt{m}.
\label{eq:app4}
\end{equation} 

For $m \le n$, we can choose $\{x_1,\dots,x_m\}$ to be
orthonormal. Consequently, 
$$\|\sum_{i=1}^m\epsilon_i x_i\|^2_2=\sum_{i=1}^m\epsilon_i^2 = m,
$$
and thus, $T_2=T_2(B_2^n) \ge 1$. On the other hand, let $m$ be an arbitrary
positive integer, and suppose that $\{x_1,\dots,x_m\}\subset B_2^n$. Then,
it is easy to show that there exists a choice of signs $\epsilon_i$,
$i=1,\dots,m$ such that
$$
\inf_{\epsilon_i=\pm1}\left\{\|\sum_{i=1}^m\epsilon_i x_i\|_2\right\}\leq \sqrt{m}.
$$
Indeed, we will show this by induction. First, note that $\|\epsilon_1
x_1\|_2=\|x_1\|_2 \le \sqrt{1}$. Next, assume that there exists
$\epsilon_1,\dots,\epsilon_{k-1}$ such that
$$
\|\sum_{i=1}^{k-1}\epsilon_i x_i\|_2 \le \sqrt{k-1}.
$$
Then (using parallelogram law), 
$$\min\{\|\sum_{i=1}^{k-1}\epsilon_i x_i + x_k\|_2^2,
\|\sum_{i=1}^{k-1}\epsilon_i x_i - x_k\|_2^2\} \le
\|\sum_{i=1}^{k-1}\epsilon_i x_i\|_2^2 + \|x_k\|_2^2 \le k.
$$
Choosing $\epsilon_k$ accordingly, we get 
$$\|\sum_{i=1}^{k}\epsilon_i x_i\|_2^2 \le k,$$
which implies that $T_2 \le 1$. Using the fact that $T_2 \ge 1$ which
we showed above, we conclude that $T_2=1$.  \qed
\end{pf}

\subsection*{{\bf Proof of Lemma \ref{lem:Litvak}}}
We now present a proof of the more general form of Lemma
\ref{lem:Litvak} as stated in \cite{Kalton} and \cite{Litvak00}
(albeit for the Banach-Mazur distance in place of $d_1$). The proof is
essentially as in \cite{Kalton}, cf. \cite{Litvak00}, which in fact
also works with the distance $d_1$ to establish an upper bound on the
Banach-Mazur distance between a $p$-convex body and a symmetric body.
{\lemma Let $0<p<1$, $q \in (1,2]$, and let $K$ be a $p$-convex
  body. Suppose that $B$ is a symmetric body with respect to the
  origin such that $conv(K) \subset B$. Then
$$d_1(K,B)\leq  C_{p,q}  [T_q(B)]^{\phi-1}[d_1(conv(K),B)]^\phi,$$ 
where $\phi = \frac{1/p-1/q}{1-1/q}$.}
\begin{pf}
  Note that $K\subset conv(K) \subset B$, and therefore $d_1(K,B)$ is
  well-defined.  Let $d = d_1(K,B)$ and $T = T_q(B)$. Thus, $(1/d)B
  \subset K \subset B$. Let $m$ be a positive integer and let $x_i, i
  \in {1,2,...,2^m}$ be a collection of points in $K$. Then, $x_i \in
  B$ and by the definition of $T$, there is a choice of signs
  $\epsilon_i$ so that $\|\sum_{i=1}^{2^m}\epsilon_i x_i\|_B \leq
  T2^{m/q}$. Since $B$ is symmetric, we can assume that $D=\{i: \
  \epsilon_i=1\}$ has $|D| > 2^{m-1}$.  Now we can write
\begin{eqnarray}
\|\sum_{i=1}^{2^m}x_i\|_K^p &=& \|\sum_{i=1}^{2^m}\epsilon_i x_i + 2\sum_{i \notin D} x_i\|_K^p 
\leq d^p \|\sum_{i=1}^{2^m}\epsilon_i x_i\|_B^p + 2^p\|\sum_{i \notin D} x_i\|_K^p\nonumber\\ &\leq& d^p T^p 2^{mp/q}+2^{mp}\alpha_{2^{m-1}}^p,
\label{eq:app1}
\end{eqnarray}
where the first inequality uses the sub-additivity of $\|\cdot\|_K$
and the fact that $(1/d)B \subset K $.  Thus by taking the supremum in
\eqref{eq:app1} over all possible $x_i$'s and dividing by $2^{mp}$, we
obtain, for any $m$,
$$\alpha_{2^m}^p \leq d^pT^p2^{mp/q - mp}+\alpha_{2^{m-1}}^p.$$ 
By applying this inequality for $m-1, m-2, ..., k$, we obtain the
following inequality for any $k\leq m$
\begin{equation}\alpha_{2^m}^p \leq d^p
  T^p\sum_{i=k+1}^{\infty}2^{-ip(1-1/q)} + \alpha_{2^k}^p \leq  d^p T^p
  \frac{2^{-kp(1-1/q)}}{p(1-1/q)\log 2} + 2^{k(1-p)}.
\label{eq:Litvak}
\end{equation}
%Note that since $\delta_K = \sup \alpha_m$, we now want to minimize
%the right hand side in \eqref{eq:Litvak} by choosing $k$
%appropriately. Since we can freely choose $m$ as large as necessary,
%we obtain the optimal value of $k$, say $k^*$, by taking the
%derivative with respect to $k$ and setting it to zero. However, $k^*$
%is not necessarily an integer. On the other hand, by choosing $k =
%k^*+1$, we can bound the right hand side of \eqref{eq:Litvak} which is
%monotonic for $k>k^*$ . This yields the following estimate for
%$\delta_K$.
Since $\delta_K = \sup_r \alpha_r$, we now want to minimize
the right hand side in \eqref{eq:Litvak} by choosing $k$ appropriately.
% Since we can freely choose $m$ as large as necessary,
%we obtain the optimal value of $k$, say $k^*$, by taking the
%derivative with respect to $k$ and setting it to zero. However, $k^*$
%is not necessarily an integer. On the other hand, by choosing $k =
%k^*+1$, we can bound the right hand side of \eqref{eq:Litvak} which is
%monotonic for $k>k^*$.  Below we present the calculations pertaining to 
%the bound on $\delta_K$. 
To that end, define $$f(k):=2^{k(1-p)} + (dT)^p\frac{2^{-k(1-1/q)p}}{p(1-1/q)\log2}$$ and 
$$A:=\frac{(dT)^p}{p(1-1/q)\log2}.$$
Since $\alpha_{2^m}^p\leq f(k)$ for any $k \in \{1,...,m-1\}$, the best bound on $\alpha_{2^m}^p$ is essentially given by $f(k^*)$, where  $f'(k^*)=0$. However, since $k^*$ is not necessarily an integer (which we require), we will instead use $f(k^*+1)\geq f(\lceil k^* \rceil ) \geq f(k^*)$ as a bound. Thus, we solve $f'(k^*)=0$ to obtain $k^*=\frac{1}{1-p/q}\log_2\left(\frac{Ap(1-1/q)}{1-p}\right)$.   By evaluating $f(k)$ at $k^*+1$, we obtain $\alpha_{2^m} \leq \left(f(k^*+1)\right)^{1/p}$ for every $m\geq k^*+1$. In other words, for every $m\geq k^*+1$ , we have 
\begin{equation}
\alpha_{2^m} \leq  (dT)^\frac{1-p}{1-p/q}\left(2^{1-p}+2^{-p(1-1/q)}\frac{1-p}{p-p/q}\right)^{1/p}\left(\frac{1}{(1-p)\log 2}\right)^\frac{1/p-1}{p(1-p/q)} \label{eq:46}.  %\forall m\geq k^*+1.
\end{equation}
On the other hand, if $m \leq k^*$, then $\alpha_{2^m}^p\leq 2^{m(1-p)}\leq 2^{(k^*+1)(1-p)}$. However, this last bound is one of the summands in the right hand side of \eqref{eq:Litvak} with $k=k^*+1$ (which we provide a bound for in \eqref{eq:46}). Consequently \eqref{eq:46} holds for all $m$. In particular, it holds for the value of $m$ which achieves the supremum of $\alpha_{2^m}$.  Since $\delta_K=\sup_r\alpha_r$, we obtain
\begin{equation}
%\delta_K \leq (dT)^{(1-p)/(1-p/q)}\left(\frac{1-p/q}{p(1-1/q)} \right)^{1/p}\frac{1}{((1-p)\log 2 )^{(1/p-1)/(1-p/q)}}.
\delta_K \leq (dT)^\frac{(1-p)}{(1-p/q)}  \left(2^{1-p}+2^{-p(1-1/q)}\frac{1-p}{p(1-1/q)} \right)^{1/p}  \left(\frac{1}{(1-p)\log 2}\right)^\frac{1/p-1}{p(1-p/q)}. \label{eq:app2}
\end{equation}
\begin{rem}
In the previous step we utilize the fact that in the derivations above we can replace every $2^m$ and $2^k$ with $m$ and $k$ respectively, thus every $m$ and $k$ with $log_2 m$ and $log_2 k$ without changing \eqref{eq:46}. This allows us to pass from the bound on $\alpha_{2^m}$ to $\delta_K=sup_r\alpha_r$ without any problems. 
\end{rem}  

Recalling the definitions of $d_1(conv(K),B)$ and $\delta_K$, note the following inclusions:
\begin{equation}
\frac{1}{\delta_K d_1(conv(K,B))}B \subset \frac{1}{\delta_K}conv(K) \subset K \subset conv(K) \subset B.
\end{equation}
Consequently $\frac{1}{\delta_K d_1(conv(K,B))}B \subset K \subset B$
and the inequality \begin{equation}d_1(K,B)=d\leq \delta_K
  d_1(conv(K),B)\label{eq:app3}\end{equation} follows from the
definition of $d_1(K,B)$. Combining \eqref{eq:app3} and
\eqref{eq:app2} we complete the proof with
%$$C_{p,q}= \left(\frac{1-p/q}{p(1-1/q)}\right)^\frac{1-p/q}{p^2(1-1/q)}\left(\frac{1}{(1-p)\ln 2}\right)^\frac{1/p-1}{p(1-1/q)} .$$
$$C_{p,q}=  \left(2^{1-p}+2^{-p(1-1/q)}\frac{1-p}{p(1-1/q)} \right)^\frac{1-p/q}{p^2(1-1/q)}\left(\frac{1}{(1-p)\log 2}\right)^\frac{1/p-1}{p(1-1/q)} .$$
\qed
\end{pf}
Finally, we choose above $B=B_2^n$ and $q=2$, recall that
$T=T_2(B_2^n)=1$ (see Proposition \ref{prop_fin}), and obtain Lemma
\ref{lem:Litvak} as a corollary with
\begin{equation}\label{eq:C_p}C(p)=\left(2^{1-p}+\frac{(1-p)2^{1-p/2}}{p}\right)^\frac{2-p}{p^2}\left(\frac{1}{(1-p)\log
2}\right)^\frac{2-2p}{p^2} .\end{equation}

\section*{Acknowledgment}
The authors would like to thank Michael Friedlander, Gilles
Hennenfent, Felix Herrmann, and Ewout Van Den Berg for many fruitful
discussions. This work was finalized during an AIM workshop. We thank
the American Institute of Mathematics for its hospitality. Moreover, R. Saab thanks Rabab Ward for her immense support. The authors also thank 
the anonymous reviewers for their constructive comments which improved the paper significantly.

\bibliographystyle{elsart-num-sort}
\bibliography{sparse}%,rayanthesis,refs}
\end{document}